\documentclass[sigconf]{acmart}

\usepackage{amsmath,amsfonts}

\usepackage{subfigure}
\usepackage{graphicx}
\usepackage{csquotes}
\usepackage{enumitem}

\def\new{\textcolor{black}} 
 
\def\AF{\textcolor{black}} 
 
\def\VF{\textcolor{black}} 

\copyrightyear{2026}
\acmYear{2026}
\setcopyright{cc}
\setcctype{by}
\acmConference[CHI '26]{Proceedings of the 2026 CHI Conference on Human Factors in Computing Systems}{April 13--17, 2026}{Barcelona, Spain}
\acmBooktitle{Proceedings of the 2026 CHI Conference on Human Factors in Computing Systems (CHI '26), April 13--17, 2026, Barcelona, Spain}
\acmPrice{}
\acmDOI{10.1145/3772318.3791465}
\acmISBN{979-8-4007-2278-3/2026/04}

\begin{document}

\date{}

\title{Helping Johnny Make Sense of Privacy Policies with LLMs}

\author{Vincent Freiberger}
\affiliation{%
  \institution{Center for Scalable Data Analytics and Artificial Intelligence (ScaDS.AI) Dresden/Leipzig, Leipzig University}
  \city{Leipzig}
  \country{Germany}
}

\email{freiberger@cs.uni-leipzig.de}

\author{Arthur Fleig}
\authornote{Equal last author contribution.}
\affiliation{%
  \institution{Center for Scalable Data Analytics and Artificial Intelligence (ScaDS.AI) Dresden/Leipzig, Leipzig University}
\city{Leipzig}
  \country{Germany}
}
\email{arthur.fleig@uni-leipzig.de}

\author{Erik Buchmann}
\authornotemark[1]
\affiliation{%
  \institution{Center for Scalable Data Analytics and Artificial Intelligence (ScaDS.AI) Dresden/Leipzig, Leipzig University}
    \city{Leipzig}
  \country{Germany}
}
\email{buchmann@informatik.uni-leipzig.de}

\begin{abstract}

Understanding and engaging with privacy policies is crucial for online privacy, yet these documents remain notoriously complex and difficult to navigate. 
We present PRISMe, an interactive browser extension that combines LLM-based policy assessment with a dashboard and customizable chat interface, enabling users to skim quick overviews or explore policy details in depth while browsing.
We conduct a user study (N=22) with participants of diverse privacy knowledge to investigate how 
users interpret the tool's explanations and how it shapes their engagement with privacy policies, identifying distinct interaction patterns. 
Participants valued the clear overviews and conversational depth, but flagged some issues, particularly adversarial robustness and hallucination risks. 
Thus, we investigate how a retrieval-augmented generation (RAG) approach can alleviate issues by re-running the chat queries from the study. %
Our findings surface design challenges as well as technical trade-offs, contributing actionable insights for developing future user-centered, trustworthy privacy policy analysis tools. %

\end{abstract}



\maketitle

\section{Introduction}
\label{sec:intro}

Almost every interaction with companies, online services, and smart devices leaves trails of personal data. 
Companies leverage techniques like hyper-personalization, powered by Artificial Intelligence (AI) and Machine Learning (ML) with real-time data sources~\cite{jain2021hyper}, to create user profiles and enable micro-targeting~\cite{chouaki2022exploring}. 
Such practices may amplify risks of manipulation~\cite{Martin2022}, automated influence~\cite{benn2022s}, and security breaches.
While companies invest in acquiring and analyzing their users' personal data, users often lack awareness of the associated privacy risks~\cite{gerber2019investigating} or have distorted perceptions of them~\cite{gerber2019johnny}. %
Privacy regulations such as the GDPR~\cite{eu2016regulation} aim to empower users by mandating transparency through privacy policies. 
However, evidence shows that many privacy policies are focused on legal compliance rather than user comprehension, targeting lawyers instead of users~\cite{schaub2017designing}. 
Since many users lack the time and expertise required to read and interpret the lengthy, technical privacy policies~\cite{obar2020biggest}, this undermines informed decision-making and reinforces asymmetries of power between companies and users.

Using Large Language Models (LLMs) to automatically assess privacy policies is a promising approach to solve this issue~\cite{sun2025empowering,rodriguez2024large,chen2025clear,zhang2025privcaptcha}.
LLMs can rephrase dense legal text into more accessible explanations, adapt to different knowledge levels, and support interactive exploration through conversational interfaces. 
These capabilities make LLMs particularly well-suited for assessing privacy policies beyond compliance checks. %
Yet, no prior work utilizes them to improve users' perceived comprehension and risk awareness by automatically assessing and interpreting privacy policies, and allowing for \textit{interactive} exploration of any website's privacy policy \textit{in-situ}, i.e., as they visit it. 

Extending preliminary results~\cite{priorWork}, we contribute the Chrome extension \textbf{PRISMe} (Privacy Risk Information Scanner for Me), to our knowledge the first to combine LLM-based privacy policy assessment with:
(i) an interactive dashboard that provides an automated high-level assessment; 
(ii) a suggestion-supported chat for deeper engagement with the policy; and 
(iii) explanations and chat responses that adapt to the user's preferences for detail and complexity. %
We build on design elements from existing transparency-enhancing tools and combine these elements %
with the adaptive language capabilities of LLMs to create a layered interaction: users can skim key assessments, verify details in a dashboard, or explore policies conversationally at a level of detail that suits their needs.
Guided by this design, we address the \textbf{research questions}: 
\begin{enumerate}[label=\textbf{RQ.\arabic*}, itemsep=0pt, leftmargin=*]
    \item\label{item:rq-understandable} 
    How do users with varying privacy knowledge levels interpret PRISMe’s privacy policy explanations?

    \item\label{item:rq-awareness} 
    How does using PRISMe shape users’ awareness of privacy risks?%
    
    \item\label{item:rq-usability} 
    How does PRISMe support users' privacy policy-related \\information-seeking and exploration, and what challenges do users encounter in everyday use?

    \item\label{item:rq-technical}
    What types of LLM errors appear in PRISMe’s responses and to what extent does a retrieval-augmented generation refinement help mitigation?

\end{enumerate}

Our focus is on designing and empirically evaluating how LLM-based privacy policy assessment, when integrated into an interactive system, can meaningfully enhance user engagement with privacy information, with user interests prioritized over legal compliance checks.
To this end, we follow a holistic approach. 
After thoroughly assessing the current landscape of transparency-enhancing tools for privacy policies, we designed PRISMe with usability for people with varying privacy knowledge levels in mind. 
We then conducted a mixed-methods user study (N=22) with a heavy qualitative focus, not to assess the impact of PRISMe on user actions outside the lab study, but to evaluate how participants interpret PRISMe’s explanations, how the tool shapes their engagement with privacy policies, and what challenges arise in practice. 
Our findings suggest that PRISMe can tremendously help users who lack online-privacy risk awareness and comprehension~\cite{gerber2019johnny}, by communicating relevant privacy protection information. PRISMe also encourages exploration of privacy policy information.
While the study revealed enthusiasm for both comprehensible summaries and conversational depth, participants raised concerns, particularly about %
adversarial robustness and hallucination risks in LLM outputs. 
Deeming such issues pressing for designing usable privacy tools, we built on these insights, systematically analyzed all LLM responses from the study, and implemented a RAG refinement into the chat. 
Re-running participants’ queries on the RAG-based backend showed that hallucinations were eliminated and adversarial robustness improved. \VF{However, some queries, i.e., chat queries going beyond the privacy policy, were now answered with: "The privacy policy does not provide sufficient context to reliably answer this question."} 
This cycle -- from user-centered design, to empirical study, to technical refinement -- constitutes the core contribution of our paper. 
Specifically,
we make the following \textbf{contributions}:
\begin{itemize}
    \item We provide an assessment of the current landscape of transparency enhancing tools for privacy policies.
    \item We present PRISMe, a browser extension that combines design elements from existing transparency tools with LLM-powered conversational interaction and that we make publicly available.\footnote{\url{https://github.com/Freiberger97/PRISMe_Privacy-Risk-Information-Scanner-for-Me}}
    \item We evaluate PRISMe through a mixed-methods user study (N=22), focusing on qualitative insights. %

    \item We identify 8 categories of issues in LLM responses (hallucinations, adversarial robustness, euphemistic language, generic responses, etc.) and demonstrate to what extent RAG alleviates these issues in follow-up evaluations.

\end{itemize}
By closing the loop between user-centered design, empirical evaluation, and system refinement, our work provides actionable insights for creating the next generation of interactive privacy policy analysis tools.

\section{Related Work}\label{sec:related}
\AF{After outlining current challenges with privacy policies in Section~\ref{sec:policy}, we provide an overview of transparency-enhancing technologies for privacy policies in Section~\ref{sec:assistants}. 
We then summarize the identified gaps and clarify PRISMe's positioning in Section~\ref{sec:gaps}.}
\subsection{\new{Challenges with Privacy Policies}}
\label{sec:policy}

Privacy policies aim to mitigate the \emph{information asymmetry} between service providers and users, empowering individuals to make informed decisions about their data~\cite{malgieri2020concept,Zaeem2020}. However, they are often designed for legal compliance, with dense, lawyer-centric language~\cite{schaub2017designing}. 
Their length, missing contextualization, and abstract phrasing add to the difficulty for users to match privacy policies with their preferences and find answers to their specific questions~\cite{windl2022automating,mhaidli2023researchers}. Legal regulations, e.g., GDPR~\cite{eu2016regulation} or the California Consumer Privacy Act~\cite{CaliforniaStateLegislature2018}, do not prevent persuasive language, which might obscure unethical practices and create a false sense of trust~\cite{pollach2005typology,belcheva2023understanding}. 
Thus, users reading and understanding privacy policies is rare~\cite{reidenberg2015disagreeable,steinfeld2016agree}, which results in informational unfairness~\cite{freiberger2024legal}.
Generative AI (by collecting usage data for model training)~\cite{lee2024deepfakes} and augmented reality (by collection of rich sensor data)~\cite{nair2023truth} complicate data management practices further and exacerbate privacy risks and transparency issues~\cite{becher2021law,belcheva2023understanding,transparency}.

\subsection{Transparency-Enhancing Technologies}
\label{sec:assistants}
To counter the informational unfairness in privacy policies, %
several transparency-enhancing technologies have been proposed. 
In contrast to \textit{semi}-automated approaches to assess privacy policies~\cite{wilson2016demystifying,wilson2018analyzing,gebauer2023human}, our tool also targets users without any expertise in privacy. 
Hence, we focus on automatic analysis of privacy policies.

\subsubsection{Reliance on a Privacy Language Representation}
Early \new{efforts relied on} %
the now-obsolete P3P privacy language~\cite{cranor2002web}. 
Privacy Bird~\cite{cranor2006user}, the first tool for automatic assessments, allows users to preconfigure privacy sensitivity levels, runs in the background while browsing, and \new{relies on static,} rule-based risk evaluations \new{to display warnings using a traffic-light system}. 
\new{However, its fixed criteria limit adaptability to the fast-changing digital landscape.
More importantly, the tool lacks a mechanism for users to query specific details about the policy and the tool's warnings, which limits its capacity to educate users or address their individual concerns.
For PRISMe, we draw inspiration from evaluating privacy policies in the background with minimal user intervention and presenting the results through a traffic-light system.}

The first privacy nutrition label~\cite{kelley2009nutrition} %
\new{visualizes key policy aspects in a \new{tabular} format, highlighting data collection types, purposes, and consent procedures.
Similar to Privacy Bird, its main limitations are static, predefined criteria and no mechanisms for users to ask questions or explore policy details interactively.
For PRISMe, we borrow from the idea of a concise, visual summary of key privacy policy aspects to enable users to quickly grasp the most critical privacy-related practices without needing to read the entire policy.}

More recently, Grünewald et al.~\cite{grunewald2023enabling} developed a layered privacy dashboard with icons and a \new{separate} RASA X-based chatbot based on their previously introduced TILT policy language~\cite{grunewald2021tilt}.
\new{
While the chatbot facilitates interactive exploration of privacy policies, its reliance on predefined conversation flows restricts its capacity for natural, dynamic dialogues and interpretation of the evaluation results.
For PRISMe, we incorporate the approach of presenting minimal information initially, allowing users to uncover additional details incrementally based on their engagement.}

\new{The overarching limitation of all the above tools is their dependence on a machine-readable privacy policy representation, which is typically unavailable.}

\subsubsection{Reliance on Processed Data}
Many tools %
rely on already processed data~\cite{guo2020poli} or crowd-sourced privacy policy annotations~\cite{Roy2024,wilson2016crowdsourcing}. 
For example, Poli-see~\cite{guo2020poli} uses icons on a circular dashboard to visualize data transmission within the provider's organization (second ring) or to third parties (third ring). %
Hovering over icons reveals data usage and consent requirements. 
ToS;DR~\cite{Roy2024} grades common web services and provides color-coded (green, yellow, and red) lists of potential issues. 
While these approaches reduce the computational complexity of real-time processing and leverage collective knowledge to identify common privacy concerns, such reliance introduces scalability and adaptability limitations.

\subsubsection{Assessment via Natural Language Processing}
\new{With the discontinuation of P3P and advances in Natural Language Processing, %
automated privacy policy assessments shifted to plain-text policies.}
In 2013, the Usable Privacy Project began automating assessments using NLP and creating corpora to support this~\cite{sadeh2013usable}. 
Of their work, the most relevant for our tool are PriBot~\cite{harkous2016pribots} and the underlying Polisis~\cite{harkous2018polisis}, which uses privacy-specific word embeddings and ML classifiers trained on the OPP-115 dataset~\cite{wilson2016creation}. 
Polisis predicts privacy icons, while PriBot was the first notable approach to inform users in a chat-based interaction about policy content. 
It decides whether the user input is a statement or a structured query before ranking which segment of the policy is the most suitable answer.
Since the returned segments are direct policy quotes, users may still struggle \new{with comprehension}. %
Our idea is to aid comprehension with clearer explanations, but still cite policy evidence (upon request).

Similarly, based on Polisis, Windl et al.~\cite{windl2022automating} \new{developed} PrivacyInjector, %
\new{which overlays icons on website elements to show context-relevant privacy policy details such as cookie information near banners, with additional details in a sidebar.}
\new{A user study showed} the tool enhances users' decisions-making, albeit without significant influence on privacy concerns. 
Users suggested reducing text length and interpreting the severity of privacy threats. %
\new{Drawing from these suggestions, our tool interprets these %
and offers various levels of text lengths.}

Dashboard-based approaches have also advanced with NLP improvements. 
PrivacyInsight~\cite{bier2016privacyinsight} \new{does not offer a comprehensive policy overview of potential issues but} focuses more narrowly on %
visualizing data transmission and their purpose, highlighting to what degree personalized data may be linked. 
\new{While this can raise privacy awareness, interpretations are left to the user.}
\new{PrivacyCheck~\cite{nokhbeh2020privacycheck} evaluates a privacy policy through 10 static questions on user control and GDPR, respectively (17 of them yes/no questions).
The respective scores are calculated by ML models.
While it determines the \enquote{market sector} of a website and displays three competitors in the same sector with better scores, %
users not satisfied with the responses to mostly binary questions cannot query specific information or ask for further or simpler explanations.}

\subsubsection{LLM-based Privacy Policy Assessment}
More recent approaches use LLMs to extract key details in a policy showing comparable performance to traditional NLP techniques~\cite{rodriguez2024large,zhang2025privcaptcha}. 
PrivCAPTCHA~\cite{zhang2025privcaptcha} extracts collected data types from a policy using a LLM before presenting the information to users as a simple CAPTCHA-like puzzle, facilitating interaction. 
Their work highlights the importance of engaging, interactive presentation of policy information. %
CLEAR~\cite{chen2025clear} limits its scope to generative AI applications, providing a contextual risk assessment for sensitive data entered by users in their prompt inputs. 
Their work shows the potential of LLMs to \textit{interpret} privacy policy information regarding potential risks induced as well as their capabilities in simplifying policy information, both requirements for PRISMe.
Hamid et al.~\cite{hamid2023genaipabench} report promising results for LLMs integrated in a chat-based privacy policy Question Answering (QA) assistant.
The potential of LLMs to increase user confidence, decrease cognitive load and enhance comprehension of privacy policies has been demonstrated by Sun et al.~\cite{sun2025empowering}. 
Their agent first sections the policy along the data practice categories introduced by Shomir et al.~\cite{wilson2016creation}, uses the LLM to summarize the policy statements in each category, and allows for question answering.
However, it is not integrated in the users' browser experience like PRISMe but requires users to find and parse the privacy policy URL in the tool manually.

\subsubsection{ML Classifiers}
\new{%
Another body of literature focuses on ML classification of legal compliance of privacy policies, targeting legal experts.}
Claudette~\cite{contissa2018claudette} uses a Support Vector Machine (SVM) to check GDPR compliance across criteria like information comprehensiveness, substantive compliance, and clarity. 
\new{Their study indicated that \enquote{none of the analyzed privacy policies meets the requirements of the GDPR}~\cite{contissa2018claudette}.}
\new{Sánchez et al.~\cite{sanchez2021automatic} introduce an SVM-driven approach that automatically assigns \enquote{positive} or \enquote{negative} tags to each policy statement, trained on manually tagged sentences from 4 privacy policies. 
It calculates scores from the (relative) amount of positive and negative tags.} 
Similarly, GDPR-completeness is classified by several methods~\cite{torre2020ai,amaral2021ai,xiang2023policychecker}. 

PrivacyGuide~\cite{tesfay2018privacyguide} uses 11 criteria, primarily based on the GDPR, with ratings (green, yellow, red) that reflect compliance (good, neutral, bad). 
It uses classifiers to identify relevant sentences from the privacy policy, which are then fed into a risk prediction engine to identify the rating. %
\new{PrivacyGuide provides direct policy quotes as evidence for its scoring, but does not provide ways to simplify and explain the given quote. 
The criteria are fixed and users cannot query specific information.}

The above approaches primarily assess GDPR compliance from a legal perspective, identifying violations such as inadequate information comprehensiveness or missing user consent mechanisms. While useful for legal experts and organizations, these tools struggle to effectively communicate and explain scores and potential risks to users without legal expertise.
For instance, a score indicating a 75\% compliance may leave users uncertain about the implications for their data. 
Moreover, the fixed criteria in these tools may overlook non-compliance issues that users deem important but are orthogonal to compliance, such as (allowed) data-sharing practices with specific third parties users do not approve of.
Hence, our focus is less on automatic assessment of (GDPR) compliance, but more on providing insights into privacy policies beyond jurisdictional constraints, allowing users to dig deeper regarding what \textit{they} deem important.

\subsubsection{Shorter and Tailored Privacy Policies}
Efforts to \new{condense} privacy policies \new{yielded} mixed results.
While \new{highlighting critical practices can boost awareness~\cite{ebert2021bolder}}, %
removing well-known facts to highlight more critical issues can reduce overall user awareness~\cite{gluck2016short}. 
Goram et al.~\cite{goram2023human} explored \new{tailoring privacy information to user preferences and concluded} %
\enquote{a long road lies ahead}.
Presenting all relevant information concisely \new{and tailored to user interests} remains the most promising approach, \new{which we thus pursue.}

\subsubsection{Beyond Privacy Policy Analysis}
\new{While we focus on analyzing raw policy text,} transparency-enhancing technologies extend beyond privacy policy analysis. 
For instance, Van Kleek et al.~\cite{van2017better} visualize smartphone app data flows based on traffic logs, and Gerber et al.~\cite{gerber2023don} explore nudging to enhance consent awareness. %
\new{Some} tools focus on %
service providers\new{, aiding compliance by user control dashboards~\cite{raschke2018designing}} %
or privacy icons for service providers to easier convey privacy-related information~\cite{holtz2011towards, MozillaPrivacyIcons2020, rossi2019dapis}. 
However, without regulatory pressure, privacy icons \new{remain underutilized.} %
We chose not to use them in our dashboard, as they would just replace single words and require familiarization.

\subsection{\new{Identified Gaps and PRISMe's Positioning}}\label{sec:gaps}
\new{Since \textit{machine-readable privacy policy representations} are not yet widespread, pioneering tools that rely on them, e.g., Privacy Bird, can provide inspirations but have limited real-world applicability. 
Similarly, tools based on \textit{processed (crowd-sourced) data} face scalability and adaptability issues. 
Approaches focused on \textit{(GDPR) compliance} can be valuable to some (e.g., legal experts), but are of limited use to non-expert users when it comes to communicating risks and addressing users’ individual preferences or concerns.
\textit{Static rules and predefined criteria} impose a one-size-fits-all solution that limits adaptation to users and evolving privacy practices.
The \textit{lack of interaction} in many tools leaves users without the ability to query specific details or understand the rationale behind evaluations, diminishing their awareness potential and engagement.
LLM-based transparency-enhancing technologies remain in their early stages. A solution that integrates into the natural browsing behavior of users and does not require additional overhead for users like manually looking for the privacy policy is still missing.
}

PRISMe bridges the identified gaps by integrating automated privacy policy analysis in-situ, with a user-centered, interactive approach.
Leveraging advancements in LLMs, PRISMe bypasses the need for machine-readable representations or pre-annotated and crowd-sourced datasets, and directly analyzes raw, plain-text privacy policies. 
This capability allows PRISMe to identify and evaluate criteria \textit{dynamically} (as opposed to static), offering \textit{adaptable explanations} (as opposed to no or static explanations) and allows \textit{user queries} on and beyond the policy text. 
PRISMe focuses on empowering users in making informed privacy decisions rather than compliance assessments based on legal frameworks. By offering layered, personalized explanations and fostering exploratory engagement, PRISMe aims to support users regardless of their background knowledge.

\section{Design Process -- Our Road to PRISMe}\label{sec:design-process} %

To develop PRISMe, we reviewed existing tools' concepts and shortcomings (Section~\ref{sec:related}). 
Analyzing literature revealed a general sense of \textit{helplessness}, a preference for seamless browsing, and \textit{low engagement with privacy decisions}. 
These insights resulted in the following \textbf{design considerations}: %

\begin{enumerate}[label=\textbf{DC.\arabic*}, itemsep=0pt, leftmargin=*]
     \item\label{item:dc-clear-communication} Communication should be clear, adaptable, and comprehensible for a wide range of users.
     \item\label{item:dc-nondisruptive} The tool should not disrupt the browsing experience too much and should offer immediate feedback.
     \item\label{item:dc-playful-exploration} There should be an exploratory, ideally \new{easy and engaging} aspect to understanding privacy policies.
     \item\label{item:dc-adaptability} The tool should adapt to different privacy information requirements and across various types of websites.
\end{enumerate}

From these considerations, we initially designed three independent LLM-based \textit{facets} of the tool, focusing on different aspects inspired by different elements of related work. 
These designs were tested within our research group \new{and then refined based on iterative feedback cycles and a pilot study}.

\textbf{\em Facet 1: Colored Scrollbar Feedback. }
This facet aimed to highlight privacy policy issues without requiring users to open the tool.
Inspired by Privacy Bird~\cite{cranor2006user}, we used colored website scrollbars (green, yellow, or red) to reflect the overall policy rating based on the assessed data protection practices. 
We wanted to raise awareness with minimal intrusion and provide immediate feedback, as suggested by Patil et al.~\cite{patil2015interrupt}. 

\textbf{\em Facet 2: Chat-Based Interactive Exploration. }
This facet centered around an interactive, chat-based interface where users could ask specific questions about the privacy policy and get answers tailored to their level of involvement. 
We drew inspiration from other privacy chatbots~\cite{harkous2016pribots,grunewald2023enabling} but utilized the flexibility of LLMs to dynamically adapt to user queries. 
This version allowed for deeper exploration requiring active user engagement, as users had to think of their own questions.

\textbf{\em Facet 3: Dynamic Dashboard Assessments. }\label{para:facet-dashboard}
We designed this facet inspired by existing dashboard-based approaches~\cite{nokhbeh2020privacycheck,tesfay2018privacyguide,bier2016privacyinsight}, which offer structured privacy assessments. 
However, instead of using fixed criteria, we implemented a dynamic assessment approach, where the assessment criteria for the privacy policies were tailored to the specific context of the website (e.g., processing health data vs.\ an email address). 
Results were displayed using smileys to communicate the severity of issues, and users could click on these smileys to get more detailed explanations. %

\begin{figure*}
   \centering
   \includegraphics[width=0.95\linewidth]{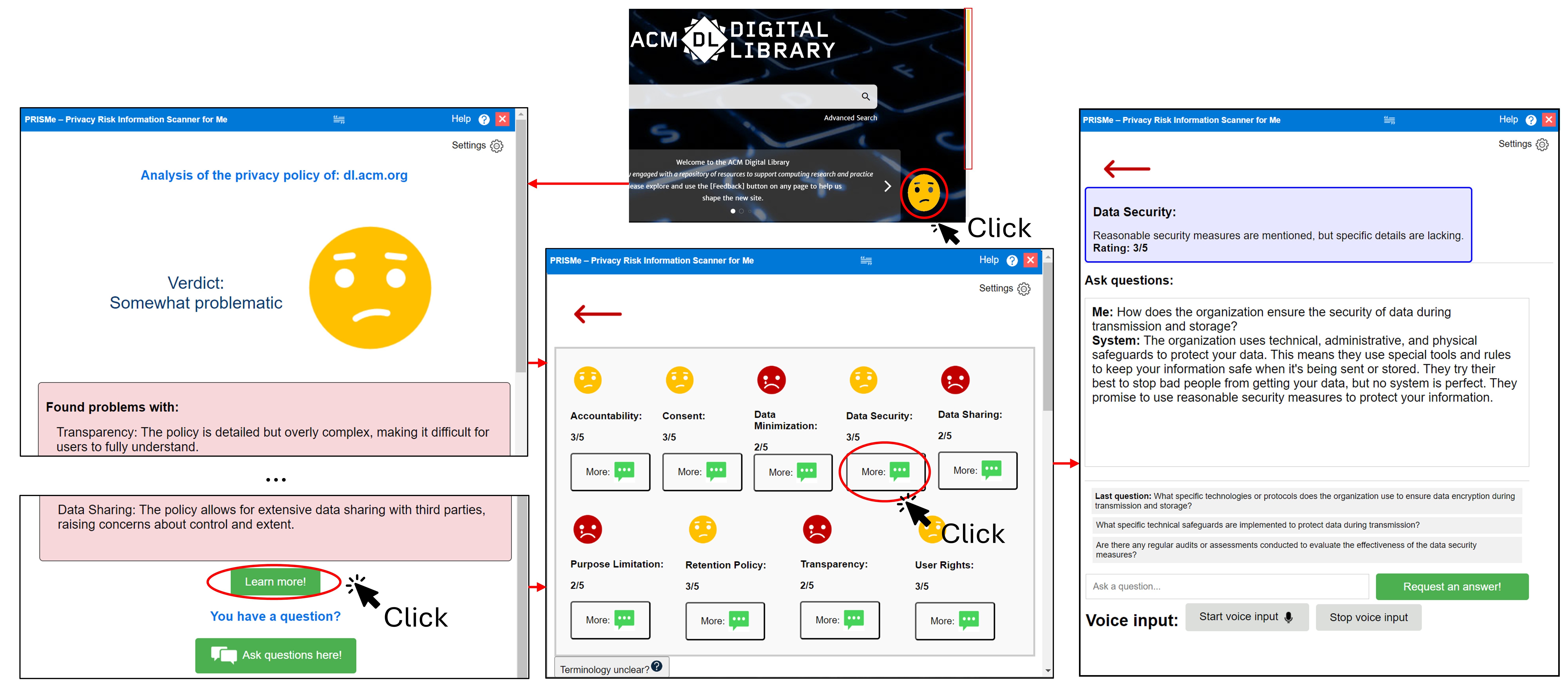}   
   \caption{Possible usage flow of PRISMe: The user first visits a website. Our tool features privacy alerts via colored scrollbars and a point-of-entry smiley icon (top middle). Clicking the smiley opens an Overview Panel (left) summarizing key privacy issues, with navigation to a Dynamic Dashboard and chat interface. The dashboard (bottom middle) provides detailed policy assessment criteria, which allows to go into a chat interface (right) by clicking on "More" below a criterion.}
   \label{fig:ui}
   \Description{The image provides a multi-step walkthrough of using Privacy Risk Information Scanner for Me on the ACM Digital Library (dl.acm.org) website to review its privacy policy. The sequence of interactions is illustrated with arrows showing clicks and actions. Here's a description of each section:
    Top Middle
    The ACM Digital Library homepage is displayed with the logo, search bar, and a yellow concerned face icon (indicating a "Somewhat problematic" privacy policy verdict).
    There is an arrow next to the smiley, showing to click on it for more information on the policy.
    Left
    This shows the PRISMe - Privacy Risk Information Scanner displaying the privacy policy's analysis for dl.acm.org, with a verdict of "Somewhat problematic." The highlighted issues are transparency and data sharing.
    An arrow points to a clickable buttons labeled "Learn more!" above another button "Ask questions here!" to get further details.
    Bottom Middle
    After clicking "Learn more," a more detailed dashboard breakdown of the policy's components is shown, including various categories such as Accountability, Consent, Data Minimization, Data Security, Data Sharing, Purpose Limitation, and others.
    Each category is rated with green, yellow or red colored smiley faces, showing satisfaction levels (e.g., Data Security is rated 2/5 with a red sad smiley). An arrow points to the "More" button under the Data Security section.
    Right
    After clicking "More" on the Data Security section, details on the site's security measures are displayed. It mentions that the policy provides "reasonable security measures" but lacks specifics, giving it a 3/5 rating.
    Below, there is a Q and A interface where the user asks, "How does the organization ensure the security of data during transmission and storage?" The system responds, explaining the use of technical, administrative, and physical safeguards to protect data, though acknowledging no system is perfect.
    The bottom of the section has a text input field for additional questions.
    In summary, the image illustrates how a user can interact with the PRISMe scanner to assess the privacy policy of the ACM Digital Library, focusing on data security and receiving additional insights through interactive features.}
\end{figure*}

\textbf{Integration into a Comprehensive Solution (PRISMe). }
Following discussions within our research group and with other HCI researchers, we combined these three facets into PRISMe to offer a layered complexity interface similar to Grünewald et al.~\cite{grunewald2023enabling}. 
With PRISMe, we wanted to allow users to interact with privacy policies in a manner that matches their level of interest and expertise, providing more in-depth information the more they explore. %

\textbf{Pilot Study. }
\new{To refine the design of PRISMe, we ran a pilot study with an early prototype and participants (4 male, 2 female) different from the main study. Based on their feedback,} we improved performance by caching LLM assessments \new{and user input, and improved the scraping of policies. We re-arranged frontend interface elements and added a plain-text display of the website's scraped policy. 
We ensured that individual user preferences on length and complexity are considered in the policy assessment right from the start.}
Finally, we added \new{speech-to-text} as an alternative to typing. %

\section{Privacy Risk Information Scanner for Me} %
\label{sec:sys}

We explain PRISMe's policy assessment (Section~\ref{subsec:policy-assessment}), user interface (Section~\ref{subsec:frontend}), and backend (Section~\ref{subsec:backend}) as used in the main user study. \AF{We will describe the post-study integration of the RAG in Section~\ref{subsec:rag}.} %

\subsection{Policy Assessment}
\label{subsec:policy-assessment}
We follow the LLM-as-a-judge paradigm~\cite{zheng2023judging} for privacy policy assessment. We instruct GPT-4o~\cite{OpenAI2024} (without fine-tuning) to dynamically identify multiple assessment criteria, 
to provide users with meaningful, context-aware information, whether they are browsing an e-commerce store or a health-related platform (\ref{item:dc-adaptability}). 
\VF{The 128k token LLM context window\footnote{https://platform.openai.com/docs/models/gpt-4o} allows us to fully ingest all privacy policies we encountered during our testing and the user study -- the longest privacy policy was 41249 tokens.} %
Each criterion is rated on a 5-point Likert scale by the LLM. 
Our prompt utilizes impersonation~\cite{salewski2024context}. \VF{We chose to let the LLM impersonate an expert in data protection who is also a member of an ethics council, to frame privacy as a balance of interests between users and service providers and to capture the real-world tensions that matter for everyday decision-making. 
In our pilot tests, this framing helped the LLM surface core privacy issues (e.g., what data is collected, sharing practices, user control) alongside transparency and fairness considerations from a consumer protection perspective, %
which aligned with PRISMe’s goals.
The prompt} guides the LLM with a multi-step reasoning process, inspired by chain-of-thought prompting~\cite{kojima2022large}. It is provided in Appendix~\ref{app:prompting}.

The overall rating we assign to a policy is calculated as the mean of all criteria ratings. 
\AF{We investigated the quality of this LLM-as-a-judge approach to ensure that the underlying LLM-generated criteria and scores were reasonable and faithful to the policy text. 
To this end, first all authors reviewed a sample of assessments and compared them against the corresponding policy passages, focusing on privacy and information fairness. 
Second, we sought external feedback from an ethicist and a jurist (not study participants) on a set of exemplary assessments to verify that the criteria and explanations were sensible from both ethical and data-protection perspectives, which both confirmed. 
Third, we utilized crowd-sourced privacy policy ratings from TOS;DR~\cite{Roy2024} to calibrate overall assessments (see Section~\ref{subsec:frontend}) using 20 benchmark websites (10 positively rated and 10 negatively rated on ToS;DR), and qualitatively checked our criteria assessments to fit the sentiment of annotations made in TOS;DR. 
Finally, for all predefined websites of the user study, we manually confirmed before the study that the assessments captured the key aspects of the respective policies.
In addition, users can delve into depth and contextualize these assessments through the chat interface (Section~\ref{subsec:frontend}) to gain a better understanding of the LLM judgments.
}

\subsection{Frontend with Multi-Layered Complexity}\label{subsec:frontend}
Our interface has \new{\textit{Point-of-Entry Signaling} and \textit{Overview}, \textit{Dashboard}, \textit{Chat}, \textit{Settings}, and \textit{Policy Text} panels}.

\textbf{Point-of-Entry Signaling. }
When users visit a website, they are automatically alerted to privacy concerns through colored scrollbars and intuitive smiley icons (Figure~\ref{fig:ui} top middle).
The smiley icon (green, yellow, or red) reflects the overall privacy policy rating (see Section~\ref{subsec:policy-assessment}) and is the first point of interaction.\footnote{red: under 2.5, yellow: 2.5-3, green: over 3; gray question mark: PRISMe cannot find/scrape the privacy policy}
Thresholds were hand-calibrated using results from the tool applied to 10 websites known for strong data protection practices like Startpage.com and 10 websites known for poor practices like TikTok.com, as identified with TOS;DR. The yellow rating threshold was set to ensure that poor data protection yields a red rating, while good practices result in a green rating. 
The color-coded smileys provide immediate feedback (\ref{item:dc-nondisruptive}) and do not require technical expertise, ensuring clear communication comprehensible for a wide range of users (\ref{item:dc-clear-communication}). 
We positioned the smiley icons on the right-hand side in the middle of the screen height, as this space is rarely covered by website elements.
However, users can drag it anywhere on the screen, giving users control of how and when they interact with the tool without being forced into unnecessary disruptions (\ref{item:dc-nondisruptive}).

\textbf{Overview Panel.}
Clicking the smiley opens the Overview Panel (Figure~\ref{fig:ui} left), summarizing critical issues (assessment criteria scored below 3) and linking to the dashboard and chat interfaces for deeper exploration.
Communicating only the most problematic aspects avoids overwhelm (\ref{item:dc-clear-communication}) and minimizes disruption (\ref{item:dc-nondisruptive}).
Links to the chat and the Dashboard Panel allow users to explore deeper without forcing them into it, supporting adaptability to different information needs from users based on their personal privacy concerns (\ref{item:dc-adaptability}).

\textbf{Dashboard Panel.}
The Dashboard Panel (Figure~\ref{fig:ui} bottom middle) provides detailed scores for each criterion, displayed as smiley icons, similar to the overall assessment.\footnote{red: under 3, yellow: 3, green: over 3} 
The given score is explained below the dashboard.
Below each criterion is a button that leads to a criteria-specific chat, inviting users to learn more in an interactive manner (\ref{item:dc-playful-exploration}).
A button below the individual ratings opens up criteria descriptions (\ref{item:dc-clear-communication}).

\textbf{Chat.}
We implemented two different chat windows to provide flexible, user-driven exploration (\ref{item:dc-playful-exploration}): one for general questions accessible through the Overview Panel \new{(General Chat)}, and one for specific criteria accessible through the Dashboard Panel \new{(Criteria Chat, Figure~\ref{fig:ui} right). They differ in their system prompt (see Appendix~\ref{app:prompting})}. 
In both, in addition to entering their own questions, users are given three dynamically generated suggestions for questions (with GPT4o-mini~\cite{o_mini}). \new{Providing suggestions was inspired by Ravichander et al.~\cite{Ravichander2021}}.
Query suggestions are generated again after each answer (which is not streamed but shown once finished) to the previous question. 
This allows user exploration in a structured way without requiring extensive effort or technical knowledge (\ref{item:dc-clear-communication}, \ref{item:dc-adaptability}). %
At the same time, we respect users who wish for a deep-dive using their own custom questions and provide voice and keyboard input options for that. 

Since chats are website-specific, an important feature is that when users visit a new website, we include the last asked question for each respective criterion (and the General Chat) from the interactions on the previous website as one of the three suggestions. 
This facilitates website comparison, e.g., asking the same question on different websites.
The chat history for each Criteria Chat and the General Chat is cached for each visited domain (but can be deleted). %
This makes it easier for users to come back to a website assessment and continue the conversation where they left off (\ref{item:dc-nondisruptive}), further facilitating website comparisons.

\textbf{Other Panels.}
The \textit{Settings Panel}, accessible via a cogwheel icon, lets users customize chat response length and policy assessments (short, medium, long) and complexity (beginner, basic, expert) to match their preferences and technical expertise (\ref{item:dc-clear-communication}, \ref{item:dc-adaptability}).
In the top bar is an icon that displays a panel with the scraped policy in plaintext. 
It helps to detect flaws in scrape results and provides users with a sense of transparency, enhancing trust calibration.
The top bar also includes a help icon where users can find the most relevant information for using PRISMe outside of lab study conditions.

\subsection{Backend}\label{subsec:backend}
The backend has four components: a privacy policy scraper, a script \new{connecting the} LLM API, a script managing the database for policy assessments, %
and an activity logger. 
The backend receives the link to the privacy policy from the frontend.
The privacy policy scraper fetches the content with Puppeteer~\cite{puppeteer} to capture dynamic JavaScript content, and removes irrelevant HTML. 
As websites display privacy policies in very different ways, the scraping might fail.
If the \new{size of the} scraped content is less than 200 words, it returns empty text, resulting in a question mark icon in the frontend (Section~\ref{subsec:frontend}). 
\VF{Valid policy text is sent to the frontend for display purposes (Section~\ref{subsec:frontend}). 
In parallel, the same policy text is sent to the LLM API by the respective backend script, which also retrieves the generated assessment or response.}
The scraper correctly retrieved 67 out of 100 pages with the most web traffic in Germany according to \cite{ahrefs}. 
Given the wide variety of policy formats and all pre-defined scenario websites in the study working reliably, this performance was sufficient for our focus on studying user interactions and the value of LLM-supported policy exploration in our prototype. %
The database script stores policy assessments in SQLite to avoid redundant evaluations. \VF{By indexing assessments using the policy text, changes in policies can be detected and trigger a new assessment. 
An outdated policy's assessment, queried by matching domain, is discarded, and the new assessment is saved. }
The activity logger tracks user activity during our study, such as feature usage (e.g., chat query suggestions) and chat histories.

\section{User Study}
\label{sec:met}

To evaluate PRISMe, %
we conducted an exploratory lab-based user study (N=22) using three custom scenarios as outlined below. 
Our primary focus is on qualitative insights drawn from interviews and participants’ in-situ reflections rather than statistical generalizability.
Questionnaire data (SUS and custom questions) and telemetry data (time spent on various panels, user settings, number of questions asked and suggestions used, mouse tracking) complement our findings. 
This approach allows us to identify key usability challenges and engagement patterns across users with varying privacy knowledge levels, laying the groundwork for future refinements and larger-scale investigations. 
Prior to the study, we confirmed with the university ethics board that our study follows our institutional and local regulations.

\subsection{Experimental Design and Procedure}\label{subsec:exp-design}
We designed three scenarios for our study participants to follow (full details with user instructions in Appendix~\ref{sec:scenario}).

\textbf{Scenario 1 \enquote{Privacy Exploration on a News Media Platform and Payment Provider}} assesses how users engage with privacy policies when using PRISMe, targeting~\ref{item:rq-understandable} %
and \ref{item:rq-awareness}. %
Media and payment services expose participants to common, yet complex online services with significant data collection. 
\VF{
We selected Focus.de, a major news media platform and Paypal, a widely used payment provider (each with millions of users), because they
are likely to be used within our participant population and represent services that many users encounter in their everyday online activities. 
Their privacy policies are typical for these industries: they exhibit high complexity, are lengthy \AF{(14760 and 41249 tokens)} and, in case of the news media platform, full of legal references. 
This makes them representative examples for evaluating how participants navigate complex privacy information. 
PRISMe rated both policies as somewhat problematic (yellow smiley), with concerns on data sharing, consent, and data minimalization for users to explore. %
This scenario} allows us to evaluate the depth of users' engagement with privacy information.

\textbf{Scenario 2 \enquote{Comparing Privacy Practices}} addresses  %
\ref{item:rq-usability}
by evaluating how users compare the privacy policies of four online bookstores offering identical pricing (\AF{including shipping}) for a hypothetical purchase.
\VF{
We selected the websites to cover variation in policy characteristics, such as differences in data-processing practices, policy length \AF{(3980 to 20174 tokens)}, and clarity, while holding the price constant.
PRISMe rated two policies as problematic (red smiley) and two as unproblematic (green smiley). 
This setup provided users with a privacy-relevant choice and allowed us to} test whether users can quickly and effectively compare privacy policies using the tool’s features. 

\textbf{Scenario 3 \enquote{Free Exploration of Websites}} examines how PRISMe supports users’ privacy concerns in personalized contexts. Participants freely explore websites of their choice, providing insights into the tool's engagement potential in real life. This contributes to \ref{item:rq-understandable}, \ref{item:rq-awareness}, and \ref{item:rq-usability}. 

The study started with participants signing a consent form and completing a questionnaire on demographics, privacy attitudes, and browsing habits.
Next, the facilitator demonstrated and explained PRISMe's functionalities. 
Afterwards, the participants explored the tool independently with suggested or self-chosen websites until they felt comfortable with PRISMe, with guidance available if needed.
\AF{They then completed the three scenarios in an average of 14 minutes for Scenario~1, 9 minutes for Scenario~2, and 7~minutes for Scenario 3, for a total average of 30 minutes.}
The facilitator encouraged the participants to think aloud, to ask questions and to voice comments, and documented them. 
Afterwards, participants filled out the SUS and our custom questionnaires (see Figure~\ref{fig:ownq}). 
The study concluded with a semi-structured interview (up to 35 minutes, 18 minutes on average).  
Participants were compensated with a 15 Euro gift card. 
The study was scheduled for one hour, but participants were free to explore further websites of their choice. 
We measured total durations from 60 to 90 minutes.

The interviews were prepared and conducted according to Myers~\cite{myers2020qualitative}.
After addressing first thoughts and impressions, the interview guide (see Appendix~\ref{tab:guide}) was divided into four focus areas evaluating \ref{item:rq-understandable}-\ref{item:rq-usability}, concluding by allowing participants to elaborate on remaining open thoughts. 
We transcribed the interviews using faster-whisper (large-v3)~\cite{systran2023fasterwhisper,openai2024whisperlargev3} that we ran locally. 
We manually checked the transcripts for accuracy and consistency. 
Utilizing thematic analysis for coding~\cite{blandford2016qualitative}, we analyzed the transcripts and comments made by participants during the scenarios facilitated by Taguette~\cite{remram44_taguette}. Initial coding was done by two researchers independently, with the goal of an aggregated and cleaned set of codes through a collaborative sense-making process. 
After initial coding, both researchers reviewed the other's codes before an in-person discussion addressing clarification (23 codes) and disagreements (15 codes). 
We dropped 3 codes, aggregated 8 codes to higher levels of abstraction, and resolved slightly different naming of codes. 
The result is an aggregated and cleaned set of 61 codes (896 coded passages, Codebook in Appendix~\ref{sec:codebook}). 
Based on all codes, both researchers identified overarching topics that codes belong to and grouped them accordingly, before aggregating their results in another discussion, leading to 6 topics. %

\subsection{Participants}
In two German cities (Leipzig and Chemnitz), we recruited 22 participants (14 identifying as male, 8 as female; age range: 18-64) via mailing lists, online message boards, public events on AI, adult education centers, and convenience sampling. The participants included researchers -- in the fields IT (3), industrial production (3), chemistry (2), and law (1) --, students at a university (4), and professionals -- in the areas IT (3), education (2), real estate (1), entertainment (1), crafts (1) and healthcare (1).
\VF{Participants’ self-reported familiarity with data protection varied considerably, from no to little prior knowledge to professional experience, including one participant who identified as a data protection officer.}
All participants stated that they rarely or never read privacy policies. 

During analysis, we observed recurring usage patterns, which we synthesized into interpretive profiles. 
We present these in the Discussion (Section~\ref{subsec:user-profiles}) to contextualize how different user types perceive and respond to PRISMe.

\subsection{\VF{Post-study Integration of RAG}}\label{subsec:rag}
Post-study, we examined the system's responses to all user queries. 
Motivated by the identified issues, we implemented and ran experiments with \textit{Retrieval-Augmented Generation (RAG)}~\cite{fan2024survey,lewis2020retrieval} on users' chat queries. 
For our RAG architecture, we utilized LlamaIndex~\cite{LlamaIndex} \VF{for semantically splitting the policy into parts (chunks)}, their VectorIndexRetriever, and the "Qwen 3 Embedding 4B" model~\cite{qwen3embedding}. 
\VF{The multilingual embedding model %
has a 32k token context window, which is sufficient for all policies we encountered due to the sematic splitting.} %
\VF{Generation is based on up to 4 retrieved chunks with a similarity score of at least 0.5.} %
The prompting (see Appendix~\ref{sec:p_rag}) instructs the RAG to ensure that the LLM's responses match with the context retrieved from the policy. If this is not the case, \AF{we programmed the RAG to tell} the user that sufficient information about the privacy policy to provide a reliable answer could not be retrieved. We also instructed the model to provide generic advice when deemed appropriate, such as privacy definitions, explanations, and common practices. Generic advice is clearly marked as not being grounded in the retrieved policy.
The RAG also considers the settings in PRISMe and the previous chat history. 

\section{Results}
\label{sec:eval}
We report and analyze the results from our user study, focusing on the interview and comments data (Section~\ref{subsec:interview}), supplemented by questionnaire data (Section~\ref{subsec:questionnaire}) \AF{and telemetry data such as the amount and types of user queries (Section~\ref{subsec:telemetry}). 
Following up on some participants' comments regarding latency and computational overhead, we quantify these in Section~\ref{subsec:loadingtimes}.
Interviews and questionnaire data} motivated the analysis of chat questions and responses, where we identify and categorize issues in LLM responses (Section~\ref{subsec:chat-history}), and report results from our subsequent attempts to mitigate these issues using Retrieval-Augmented Generation (Section~\ref{sec:rag}).

\subsection{Interview \& Comments Data}\label{subsec:interview}
We identified 61 distinct codes in 6 topic areas, which we summarize below. The full codebook is in Appendix~\ref{sec:codebook}. 

\subsubsection{\textbf{Topic 1: User Attitudes, Motivations, and Behavior} (\ref{item:rq-awareness}, \ref{item:rq-usability}).}
Participants often displayed \textit{indifference} (P1, P6, P8), \textit{insecurity} (P4, P14) or \textit{misconceptions} about privacy risks (e.g., believing private browsing prevents tracking) (P5, P9) and expressed distrust in websites' data protection practices (P1, P2, P8, P9, P11, P12).

While participants see challenges in changing habits (P1, P4, P5, P6, P10, P18) and deem privacy-unrelated factors (more) important (P3, P8, P11, P12, P14, P16, P20, P22), they also expressed a strong interest in using the extension in their daily lives: \textit{\enquote{If it was available, I think I would directly install it}} (P21), \textit{\enquote{I think it would be great progress if this became standard practice}} (P20). 
Some highlighted professional applications beyond private use, such as data protection training (P18), improving website practices (P20), documentation (P20), and assessing business partners (P6, P22).

\subsubsection{\textbf{Topic 2: Information Quality and Clarity} (\ref{item:rq-understandable}, %
\ref{item:rq-usability})}
\label{sec:info_quality_interviews}
Participants praised PRISMe for \textit{simplifying} complex privacy policy language, acting as a \enquote{translator} (P20) that made data protection accessible. 
P19 noted, \textit{\enquote{You don’t need a university degree to comprehend data protection this way}}. 

While some participants noted a lack of transparency in how ratings were determined (P1, P10, P12, P17, P18, P22) and expressed confusion over the rationale behind certain assessments, \textit{\enquote{I couldn’t really understand why this lower rating was given. Or why the system came to this exact conclusion}} (P14), all agreed the simplified language significantly improved their (perceived) understanding.
All but two participants highlighted the quick, clear overview of privacy policies, enabling access to essential information with minimal effort. For example, P21 was happy with the provided overview: \textit{\enquote{You don’t have to click at all to get a basic assessment straight away}}.

Feedback on the tool’s \textit{detail levels} was mixed. Some participants found it confusing when a green overall smiley was accompanied by an alert on the Overview Panel (P8, P11, P12, P18). 
Moreover, participants desired more straight signals, such as, \enquote{This is dangerous} (P10). Others acknowledged the balance between detail and usability, with P22 stating, \textit{\enquote{I would have liked it to be more specific, but then again you have to read more}}. 
Participants found the information comprehensive and relevant. P2 summarized, \textit{\enquote{Everything important in a privacy policy is included: what data is collected, transparency, legal basis, purposes...}}.

Participants appreciated the tool’s speed and accuracy, noting they could quickly obtain desired information without many follow-up questions (e.g., \textit{\enquote{I could get what I wanted very quickly}} (P20), \textit{\enquote{a 9 out of 10}} regarding efficiency (P14)). The chat also anticipated user needs, as one participant mentioned, \textit{\enquote{I really like that it starts explaining how I can view collected data right away}} (P9), and responses were seen as well-balanced (P12).

The \textit{chat functionality} was well-received for its flexibility, handling typos, multiple languages, and sophisticated or nuanced queries (P1, P7, P15). Responses were consistent and conversational, often anticipating user needs (P9). As P22 noted, \textit{\enquote{I was surprised by how exact it answered with very specific information}}.

\subsubsection{\textbf{Topic 3: User Experience and Interface Interaction} (%
\ref{item:rq-usability}).}
\label{sec:experience_interviews}
Most participants liked the smiley icons as \textit{visual cues} providing quick, intuitive insights. P9 stated, \textit{\enquote{The smiley is great because it doesn’t interfere but [quickly shows]: Is this [privacy policy] good or not?}} Few participants found the visual cues to be intrusive and emotionally loaded, suggesting alternative symbols like exclamation marks for problematic policies (P6, P7, P10).

The tool’s ease of use was widely appreciated, described as \textit{\enquote{user-friendly, clear, and courteous}} (P6). Some participants complained about having to scroll using the extension (P8, P20). Others suggested clearer chat formatting (e.g., structured paragraphs and key highlights) (P1, P8, P10, P11, P13, P14, P18, P21). Voice input was used and valued by some, and adding audio output for accessibility was suggested (P5, P12). Some participants would have liked a faster initial assessment of a website (P1, P2, P6, P12, P15, P20), whereas the chat was perceived as quick and responsive (P3, P5, P6, P7, P8, P14, P17, P22).

The query suggestions in the chat were seen as a helpful starting point and helped participants organize their thoughts or get inspiration for their own questions (P1, P3, P9, P13, P14, P15). P4 mentioned that it also helped to quickly contextualize the information provided on the dashboard. Few participants wanted suggestions to be more specific (P3), cover a more diverse range of topics (P3), and be shorter (P6). P14 suggested a hierarchical mapping of keywords on the topic to inspire one's own questions instead of the query suggestions.

\subsubsection{\textbf{Topic 4: Tool Reliability and Trustworthiness}  (\ref{item:rq-usability})}
While participants generally appreciated the tool, concerns about reliability arose. P7 suspected hallucinated or speculative information, while P8 questioned the source of the information: \textit{\enquote{Sometimes I doubted whether the given information was from the policy itself. It was sometimes expressed more like an assumption}}.
Participants suggested adding links to the relevant policy sections (P2, P8, P10, P13, P18, P22). Concerns were also raised about the dynamic assessment criteria, which some felt lacked weighting by importance, reducing comparability between policies and perceived reliability (P3, P9, P12, P17, P22).
Some participants mentioned the scraper not being able to fetch the privacy policy on all websites in the free exploration scenario (P1, P5, P15, P22), in which case they quickly switched to another website.

\subsubsection{\textbf{Topic 5: Recommendations for Features and Functional Improvements} (\ref{item:rq-usability}).}
Participants wished for a design that goes beyond a website's privacy policy, such as suggesting alternative websites (P1, P4, P7, P12) and providing contextual feedback on current privacy settings (P15, P18). Some were open to allowing PRISMe to adjust browser privacy settings and taking them into account in the website assessment %
(P7, P8, P15).
Furthermore, participants wished for an adjustable window size (P2, P3, P8, P12, P15, P16, P21), and getting a side-by-side comparison between pages (P2, P4, P9, P12, P14, P18).

\subsubsection{\textbf{Topic 6: Impact on Users} (\ref{item:rq-awareness}).} %
\label{sec:impact_interviews}
Participants reported the tool generally raised their \textit{privacy awareness} and reminded them or helped them discover what is important to them in data protection. 
P12 commented, \textit{\enquote{I think it helped me to be a bit more sensitive to the topic of data protection. I think this short session was already useful}}. 
PRISMe sparked emotional responses, curiosity, and a new intent to pay closer attention (P1, P9).
For instance, P1 stated, \textit{\enquote{a sad face like that does something to me emotionally}} and reflected on the importance of data protection, saying, \textit{\enquote{Looking at privacy policies has never been relevant in my life before, even though that's actually a bit stupid. And I think I first need to become more aware that data protection is actually important and that websites have different levels of privacy policies}}. 

Participants were motivated to explore further, with P15 noting, \textit{\enquote{It was interesting, so I always wanted to try more}}. 
The tool also encouraged critical thinking about data protection. P11 mentioned, \textit{\enquote{I thought about what data protection problems could actually occur.}}
Participants reported improved \textit{understanding} of privacy issues, with P4 highlighting the value of explanatory sentences for assessment criteria. This newfound awareness prompted some to consider being more cautious with their data: \textit{\enquote{I can imagine that I would be more careful about who I entrust with my data}} (P12).

\subsection{Questionnaire Data}\label{subsec:questionnaire}
\begin{figure}
   \centering
   \includegraphics[width=0.75\linewidth]{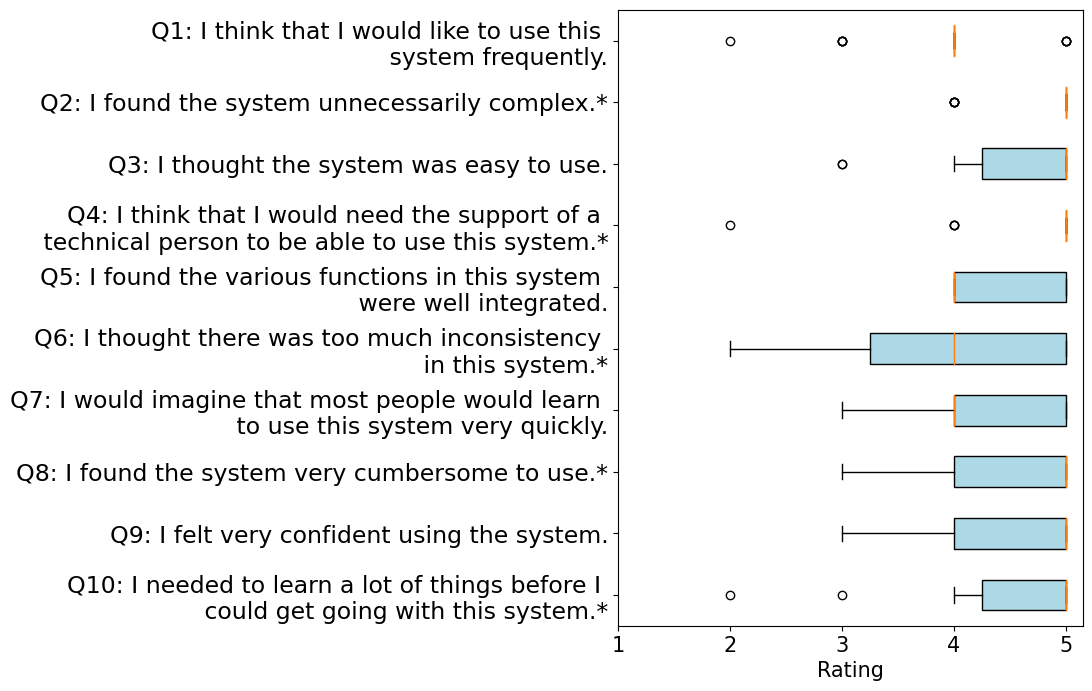}   
   \caption{System Usability Scale results (questions with * have inverted scores; higher values are always better)}
   \label{fig:sus}
   \Description{The plot is a boxplot visualization of the adjusted System Usability Scale (SUS) questionnaire results, where higher scores are always better. It displays ratings for 10 questions related to system usability, with the following observations:
- Q1: The ratings range from 2 to 5, with a median at rating 4.
- Q2*: The ratings mostly span from 4 to 5, with a median of 5.
- Q3: The ratings are between 3 and 5, with a median of 5. Outliers at 3.
- Q4*: The ratings have a median of 5, outlier at 2.
- Q5: The ratings span from about 4 to 5, with a median of 4.
- Q6*: The ratings range from 2 to 5 with a median of 4.
- Q7: The ratings fall between 3 and 5 with a median of 4.
- Q8*: The ratings span from 3 to 5, with a median of 5.
- Q9: The ratings vary from 3 to 5 with a median of 5.
- Q10*: The ratings range from about 2 to 5, with a median of 5 and
an outlier at 2.
The asterisks (*) next to certain questions denote that these questions were reverse-scored in the original SUS.
}
\end{figure}
Figure~\ref{fig:sus} shows the results of the SUS questionnaire, with scores adjusted so that a 5 represents the best possible rating. 
The lowest-scoring question relates to system inconsistencies. 
Notably, most participants did not find our tool complex to use and would use it frequently, without needing support.
The overall average score is 88.9 out of 100, well above the typical 68-70 reported in literature~\cite{brooke1996sus, brooke2013susretrospective}, placing it in the \enquote{excellent} range and indicating a high level of user satisfaction.

\begin{figure}
   \centering
   \includegraphics[width=0.75\linewidth]{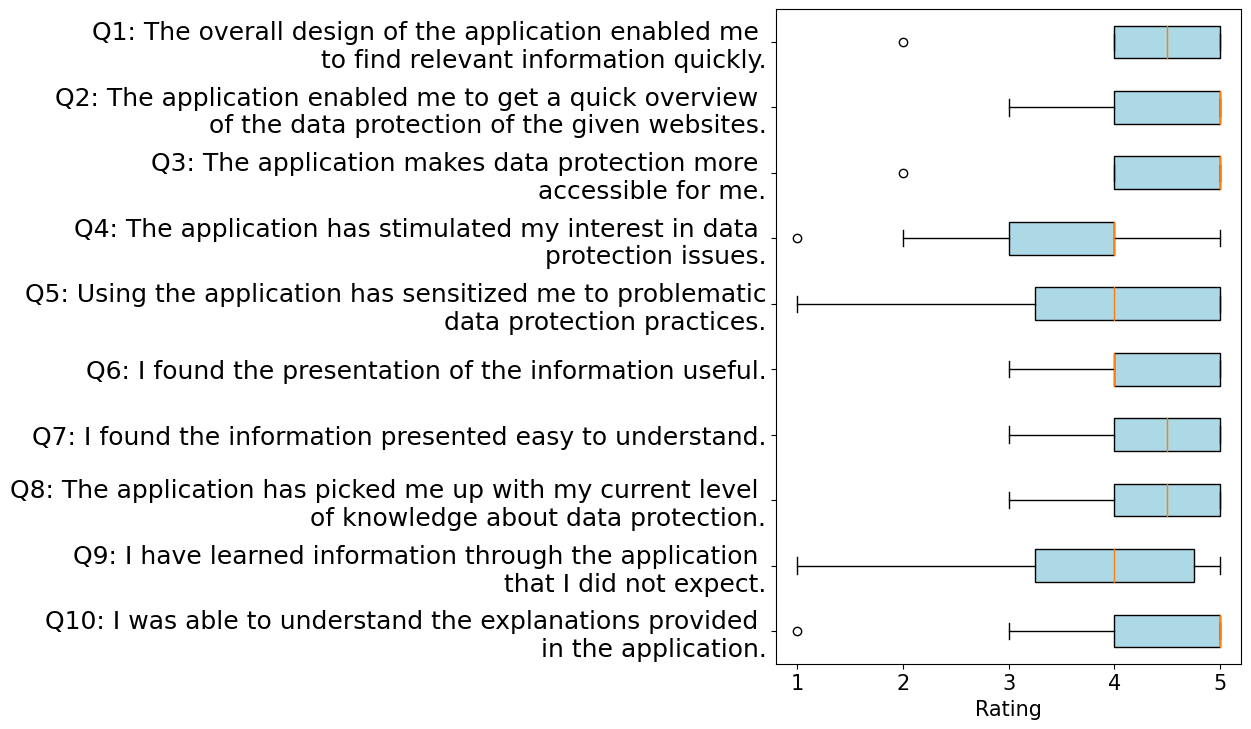}   
   \caption{Results of our questions on a 5-point Likert scale (1: strongly disagree; 5: strongly agree).}
   \label{fig:ownq}
   \Description{The image displays a horizontal box plot that evaluates responses to a specific questionnaire. Each box plot corresponds to a different question in the questionnaire. The questions are listed on the y-axis, and the x-axis represents the rating on a scale of 1 to 5.
- All questions have medians in the higher rating range (between 4 or 5).
- The questions Q5 and Q9 have a wide distribution of responses.
- There are some outliers present, especially for Q1, Q3 and Q10.}
\end{figure}

Figure~\ref{fig:ownq} illustrates participants' responses to our custom 5-point Likert scale questions related to awareness (Q4, Q5, Q9), perceived understandability (Q3, Q6, Q7, Q8, Q10), and efficiency (Q1, Q2). 
Awareness was generally raised, though for some more than for others. 
Participants (strongly) agreed on understandability-related questions, in particular PRISMe making the information accessible (Q3), it picking users up at their respective knowledge level (Q8), and comprehending the provided explanations (Q10). 
Efficiency regarding both single elements of PRISMe (Q2) and the tool as a whole (Q1) was rated particularly high. 

\subsection{Telemetry Data}\label{subsec:telemetry}

In terms of interface use, most participants concentrated on the chat panel, while a few focused on the overview and dashboard panels.
Mouse tracking data supports this: some participants immediately navigated to their specific criteria of interest to ask questions, while others explored the dashboard before occasionally asking questions.
Participants collectively asked 368 questions in the chat, with strong variation across individuals (from 4 by P16 to 46 by P1). 
Some relied entirely on suggested queries, while others posed self-formulated questions almost exclusively. 

\VF{
To better understand participants’ information needs, we analyzed and grouped the questions they entered. 
Participants often used PRISMe's assessment criteria as a starting point, to ask about issues highlighted by the tool first.
However, they also raised a range of additional concerns that extended beyond PRISMe’s categories. 
These included questions about profiling and targeting (e.g., whether a particular news site personalizes article recommendations based on user profiling), communication and contact (e.g., how to exercise their right to be forgotten in practice), compliance mechanisms (e.g., whether the service undergoes security audits to detect vulnerabilities), and general characteristics distinct from PRISMe's evaluation criteria (e.g., the availability of viable alternative services). 
Table~\ref{tab:question-categories} summarizes the thematic categories of questions participants
asked in the study and provides examples of user-initiated questions that were \emph{not} suggested by PRISMe.} %
Several participants repeatedly adjusted the length setting, tailoring responses to their preferred level of detail.

\begin{table*}[h]
\centering
\caption{\VF{Categories of questions participants asked PRISMe during the user study, with examples phrased by participants}}%
\label{tab:question-categories}
\renewcommand*{\arraystretch}{1.15}
\begin{tabular}{|p{7.25cm}|p{7cm}|}

\hline
\textbf{Category}  & \textbf{Exemplary User Question} \\ 
\hline
\textbf{Data Collection} (what data is collected and why) & \textit{What non-essential data is stored?} \\
\hline
\textbf{Purpose Limitation} (how transparent and justified purposes are) & \textit{What are the purposes for which my data will be used, and where would it need to be formulated more precisely?} \\
\hline
\textbf{Data Security} (technical and organizational safeguards) & \textit{How often is the security of the system checked?} \\
\hline
\textbf{Compliance Mechanisms} (audits, monitoring, documentation) & \textit{How is compliance with the purpose limitation documented and monitored in the privacy policy?} \\
\hline
\textbf{Data Sharing} (third parties, transfers, international processing) & \textit{Is it specified more precisely to which non-EEA countries data is transferred?} \\
\hline
\textbf{User Rights} (rights and how to exercise them) & \textit{What is the right to data portability? Which rights are unclear?} \\
\hline
\textbf{Profiling and Targeting} (implications of tracking, analytics, personalization) & \textit{What are the disadvantages for the website if it does not use any Google Analytics functions?} \\
\hline
\textbf{Consent Mechanisms} (consent characteristics and processes, policy changes) & \textit{Did I already consent?} \\
\hline
\textbf{Contact and Communication} (who to contact and how) & \textit{Is there a form for GDPR information requests?} \\
\hline
\textbf{Retention} (how long data is stored, deletion practices) & \textit{After what time are YouTube search queries deleted?} \\
\hline
\textbf{Transparency} (policy clarity, specificity, accessibility) & \textit{What measures could be taken to increase the transparency of the privacy policy?} \\
\hline
\textbf{General \& Meta Questions} (overall safety, recommendations, interpretations, alternatives) & \textit{What is the most important thing I should consider not doing when I use this website?} \\

\hline
\end{tabular}
\end{table*}

\subsection{\VF{Latency and Computational Overhead}}\label{subsec:loadingtimes}
\VF{
As noted qualitatively in Section~\ref{subsec:interview}, some participants mentioned the initial assessments loading somewhat slowly, especially during free exploration in Scenario~3, while the chat interaction was described as quick and responsive. 
To quantify these impressions, we conducted a post-study latency evaluation using 10 policies (ranging from 1727 to 41249 tokens) from our study. 
We assessed latency for initial policy assessments and chat responses, repeating every measurement three times. 
Without a pre-existing assessment in the database, assessing a privacy policy and displaying the results took 19.94 s on average (SD=1.22 s). 
With an assessment available in the database (which was the case in scenarios~1 and~2), loading times went down to 3.97 s on average (SD=0.56 s).
Most of this delay stemmed from re-fetching the website's privacy policy and checking it against the stored version (a safety step to detect policy changes before using the cached assessment).
To evaluate chat responsiveness, we selected an exemplary user question for each of PRISMe's length settings (short, medium, long). 
The average response times were 1.87 s (SD=0.56 s), 3.73 s (SD=1.20~s), and 8.32 s (SD=3.20~s) for short, medium, and long settings, respectively.}

\subsection{Issues in LLM Responses}\label{subsec:chat-history}
We analyzed all 368 LLM responses to chat queries, identifying 2 system-related and 6 LLM-related problems. 

Regarding \textbf{problems related to our system}, 
four policies during free exploration (Scenario 3) were \textbf{partially scraped}, which led to \VF{13} 
incomplete responses, e.g., \textit{\enquote{[...]There are no clear details about the security measures}}. The LLM avoided hallucinating details, offering abstract responses based on available content.
Four cases involved \textbf{context-related limitations} due to reliance on plain text rather than HTML; one missed a hyperlink to a requested form (\textit{\enquote{The specific link for the [...] online data protection request form is not provided directly in the privacy policy. [...]}}), and three failed to provide step-by-step privacy setting instructions.

As for \textbf{problems related to the LLM},
\VF{5 instances of }\textbf{hallucinations} involved fabricated information about its own assessment, about audit details not specified in the policy (\textit{\enquote{Yes, PayPal undergoes external audits and reviews to ensure that it complies with its anonymization and aggregation policies.}}), and recommending other web services without checking them (\textit{\enquote{For a news website that respects your privacy more, you might consider using [...]}}). 
\AF{
When the policy lacked the requested information, we did not classify the LLM’s use of general background knowledge as hallucination, as long as the statements were factually correct and not contradicted by the policy. 
In these cases, the LLM responded based on general knowledge rather than grounding its answer in the policy text, though without explicitly indicating this to the user.}

The LLM mirrored \textbf{overly euphemistic}, marketing-like language from policies \VF{12 times}, 
adding reassuring but uncritical conclusions that irritated some participants (P9, P12, P17).
For example, to the question, \textit{\enquote{Where will my data be transmitted to?}}, it responded, \textit{\enquote{[...] To summarize: Your data is mainly shared within TikTok and with trusted partners to operate and improve the app.}}

We found \VF{12} cases where the LLM partially ignored or \textbf{omitted relevant information} in its chat responses, unless explicitly asked. 
An example question is, \textit{\enquote{[...] and how is this communicated in the privacy policy?}}, to which the LLM responded, \textit{\enquote{[...] This is explained in the privacy policy under the relevant sections.}}
This led to user frustration, particularly among participants seeking specific details (P1, P3, P5, P17).

\VF{Eight} overly \textbf{generic responses} lacked specificity to the given policy and were perceived as less useful as they required additional prompts for clarification (P3, P9, P17).
For example, one participant asked, \textit{\enquote{What options do users have to withdraw their consent at any time and how is this process communicated?}}. 
The response was, \textit{\enquote{Users can withdraw their consent at any time by changing certain settings in their account or by contacting customer service. This process is explained in the privacy policy. [...]}}.

\textbf{Misunderstandings of user queries} were evident in \VF{4} responses. 
One participant requested, \textit{\enquote{Please show me the entirety of the categories to evaluate privacy policies.}}, to which the LLM responded, \textit{\enquote{Of course, I will summarize the most important categories and aspects of the privacy policy [...]}}.
While interrupting the flow of interaction, they were typically easy for users to identify and fix with follow-up queries (P8, P12, P16).

Two chat responses confused users by presenting seemingly contradictory information, sending \textbf{mixed signals}. 
Asked whether only the most necessary data is collected, the LLM responded, \textit{\enquote{No, PayPal collects a variety of data categories, including sensitive data. Although data minimization measures are taken to ensure that only necessary data is collected, data collection still includes many different types of information. [...]}}.
The phrasing caused unnecessary doubt or hesitation (P11, P13).

\subsection{The Effects of Adding RAG to the Chat}
\label{sec:rag}
To mitigate the identified issues with LLM responses (Section~\ref{subsec:chat-history}), we augmented PRISMe's chat with a RAG and replayed participants’ queries against this new setup. 
\VF{
To align with the RAG objectives of maximizing factual accuracy and source transparency, %
our RAG analysis examines how well each query can be supported by retrieved policy evidence, distinguishing cases with clear support, cases where the policy is too vague to justify a reliable answer, and cases where no relevant policy information exists. To this end, we manually checked all RAG responses.
}

From the participant's 368 chat queries we excluded 29 follow-ups that referenced chat history. 
Out of the 339 replicable queries, in \AF{143} cases the \textbf{RAG stated it lacks sufficient information for a reliable response} because \AF{we instructed the RAG to do so when} its retriever did not surface sufficient policy evidence. %
We manually inspected these queries: 45 were rightfully not answered, as participants asked about information outside the policy. For instance, the query, \textit{\enquote{Who are trusted partners?}} caused the LLM to hallucinate in the user study, as the respective privacy policy did not define trusted partners. Our RAG now correctly avoided answering. 
In 42 cases, it was acceptable that PRISMe did not answer, because the policy was vague or only implicit. A typical example is a question how a company ensures that only necessary data for the respective purposes is collected. Since the policy did not detail a specific process, the RAG stated that it cannot reliably answer. 
\AF{Another 13 cases stem from \textbf{partially scraped} policies.}
The remaining 43 non-answers were retrieval misses or requests requiring a holistic interpretation of the policy. 
For instance, a question on data security measures was not answered, as the retrieved chunks did not contain the relevant quote from the policy:\textit{\enquote{firewalls, data encryption, and access controls}}. 

Of the remaining \AF{196} queries, 185 %
were \textbf{correct and fully faithful} to the policy. 
Importantly, %
we observed improvements in specificity and practical usefulness: in 40 cases RAG produced noticeably more precise, actionable answers tied to explicit policy passages (for example, replacing generic advice about third-party sharing with explanations of policy-specific clauses and citations).

\AF{We report problem types identified in Section~\ref{subsec:chat-history}, comparing the issues in RAG responses with those in the original LLM responses. 
Of the 196 responses given by the RAG, none contained \textbf{hallucinations} (down from 5 in the original LLM responses in Section~\ref{subsec:chat-history}), \textbf{overly euphemistic language} (down from 12), or \textbf{misunderstanding of user queries} (down from 4). 
In all cases of incomplete responses due to \textbf{partially scraped} policies, the RAG stated it lacked sufficient information for a reliable response.} 
\AF{Eleven answers we consider problematic distribute as follows.
One of the four original \textbf{context-related limitations} remained problematic since the response was given incorrectly due to the policy not being provided as HTML.}
\textbf{Omission of information} became less frequent \VF{(5 cases, down from 12)}, %
\AF{in which case} the retriever did not return all user rights or third parties stated in the policy. 
\textbf{Generic responses} decreased \VF{to 2 instances (down from 8)}.
The problem with \textbf{mixed signals} \AF{(2 instances, no change)} persisted in the form of inconsistencies between the part of the response based on the retrieved chunks of the privacy policy and the \AF{now clearly marked} generic advice we additionally allowed to be made \AF{(see Section~\ref{subsec:rag})}. 
\AF{Finally, one response fell outside the previous taxonomy: a factual mistake caused by an incorrect LLM inference despite correct retrieval.}

\section{Discussion}\label{sec:disc}

We structure our discussion according to four interpretive profiles, which we derived from participant comments', telemetry data, and the interviews.
After introducing them in Section~\ref{subsec:user-profiles}, we discuss our findings regarding our research questions (\ref{item:rq-understandable}-\ref{item:rq-technical}) in Sections~\ref{subsec:disc-rq1}-\ref{subsec:llm-limitations}, elaborating how the different profiles are affected, and deriving design implications for future tools like ours. 
Finally, we reflect on limitations and further considerations in Section~\ref{subsec:limitations}.

\subsection{Interpretive Profiles}\label{subsec:user-profiles}
We observed recurring patterns of interaction that we synthesized into four interpretive profiles.
While these profiles are neither exhaustive nor generalizable beyond our sample, they help contextualize how different users engaged with PRISMe. 
Some overlap with existing archetypes from
Dupree et al.~\cite{dupree2016privacy} or Hrynenko and Cavallaro~\cite{hrynenko2025identifying}, while others do not match exactly, notably our \textit{Novice Explorers}, who strive to discover and learn, rather than simply disengaging.

{\textit{Targeted Explorers}} (P2, P7, P12, P17, P18, P20, P22) engaged deeply, seeking detailed and specific information. They tended to have prior privacy knowledge and clear goals, used advanced customization options, and requested evidence. They navigated the tool confidently, with a focus on asking their own questions. 

\textit{Novice Explorers} (P4, P13, P14, P19), with limited prior knowledge and confidence in understanding privacy policies, explored the tool by 
discovering and defining their informational goals. %
They used guiding-focuses features such as chat suggestions extensively and spent more time reading the initial assessment provided by the tool. %

\textit{Balanced Explorers} (P1, P3, P8, P9, P10, P11, P15, P21) combined discovery and seeking of specific information. They tended to chat about many different assessment criteria, often starting with a suggested query before going into depth with their own questions. They explored multiple features without fixating on specific elements, hence benefiting from flexible use and a broad assessment scope.

\new{\textit{Minimalistic Users}} (P5, P6, P16) prioritized efficiency, engaging minimally using the chat.
They usually stopped after a few interactions, and were often satisfied already with high-level summaries and smiley ratings, looking for concise assessments and quick insights.

\subsection{\ref{item:rq-understandable}: How do users with varying privacy knowledge levels interpret PRISMe’s privacy policy explanations?}
\label{subsec:disc-rq1}
A key insight is that PRISMe \textit{simplifies complex privacy policy language}, which all participants -- even \textit{Targeted Explorers} -- identified as crucial in enhancing users’ perceived understanding of privacy policies. 
\textit{Novice Explorers} in particular see it as a \textit{translator from legalese to plain language}, which is in line with prior research~\cite{sun2025empowering}.
The easier-to-comprehend representation encourages an exploration process, and explanations and interpretations beyond policy quotes~\cite{harkous2018polisis,tesfay2018privacyguide,windl2022automating} help users discover what is relevant to them regarding data protection.
PRISMe also overlooks spelling mistakes and adapts to different languages, thus lowering entry barriers for user queries. 
Its voice input facilitates inclusion, and participants appreciated the different modes of interaction. 
To further enhance understanding of PRISMe's explanations, future work could improve the \textit{response formatting}, e.g., clearer visual structuring and keyword highlighting. 

PRISMe's adjustable settings span from simple explanations liked by \textit{Novice Explorers} to specific details in technical terminology liked by \textit{Targeted Explorers}. 
\textit{Personalization} and high-quality suggestions are essential to avoid discriminatory outcomes, as the cognitive demand of asking the right questions or finding suitable settings may prefer high-literacy individuals~\cite{Ravichander2021}.
Hence, it should be a major goal for tools like ours. 
Potential next steps include automated customization, which has been called for~\cite{Ravichander2021,goram2023human}, and individual policy assessment criteria users can define to decrease metacognitive demands~\cite{tankelevitch2024metacognitive}. 

Our analysis of PRIMe's chat answers revealed that the used LLM sometimes sends mixed signals, making it hard for users to quickly and clearly comprehend the provided information.
This may interrupt the exploration process of \textit{Novice Explorers}, and \textit{Minimalistic Users} may lose trust in the tool. 
\AF{This motivated the implementation of the RAG.}
With RAG restricted to policy text, mixed signals disappeared but general advice (which participants queried and liked) was also lost. 
\AF{Moreover, without RAG, the LLM sometimes produced responses that were not grounded in the privacy policy but were factually correct. 
With RAG, such speculative “helpful hallucinations” vanished and instead, the RAG stated that it cannot reliably answer that query. 
While non-answering may improve trustworthiness, it also constrains exploration for \textit{Novice Explorers} and may frustrate \textit{Minimalistic Users} seeking quick answers.} 
Hence, we allowed \AF{the RAG} to supplement the (grounded) answer with generic advice, \AF{and re-ran all user queries with this setup (Section~\ref{sec:rag}),} which reintroduced two instances of mixed signals. 
This highlights a core design trade-off: strict grounding ensures fidelity, while generic advice better supports broader user needs. 
We recommend future tools to clearly signal which parts of a response are policy-based versus general knowledge.
\AF{To better navigate the trade-off, future studies could measure users’ confidence in PRISMe’s outputs, with and without strict grounding.} %

\subsection{\ref{item:rq-awareness}: 
How does using PRISMe shape users’ awareness of privacy risks?}\label{subsec:disc-rq3}
PRISMe raised participants' self-reported awareness across all profiles, with \textit{Targeted Explorers} showing the least change, which we attribute to their pre-existing knowledge. 
\textit{Novice Explorers} and \new{\textit{Minimalistic Users}} gain an improved understanding, countering privacy misconceptions. 
PRISMe encourages reflection especially for \textit{Balanced Explorers}. 
It prompts them to engage with privacy issues more deeply by reminding about them and triggering emotional responses.
These findings suggest that awareness effects are not uniform but mediated by users’ prior knowledge and interaction style.
Future work could investigate to what extent our always-displayed initial smiley raises awareness or dulls it over longer time periods. 
\VF{%
A longitudinal study would help clarify how much of the observed awareness increase is attributable to the lab setting and whether such effects are sustained. 
This could be measured by multiple awareness tests over the duration of such a study. 
Since privacy was a primary task in our user study but is typically a secondary concern in everyday browsing, real-world awareness effects may be weaker.
Evaluating PRISMe during naturalistic browsing routines would therefore provide more ecologically valid insights.
}

A potential risk is over-reliance on PRISMe’s assessments, which mitigates awareness by users feeling protected~\cite{schaub2016watching}.
The convincing nature of LLM outputs may mislead users into making misguided judgments~\cite{krugel2023chatgpt} and reduce critical thinking~\cite{lee2025criticalgenai}.
To counter this, tools should manage expectations by signaling uncertainty~\cite{kim2024m} and clearly distinguishing between grounded policy content and generic advice.

We also note that policies themselves can contain persuasive language, which can impact our initial assessment and result in uncritical and positively framed chat answers. 
Our RAG-based refinement prevented such affirmative language from being reproduced in responses, suggesting a promising approach. 
We discuss this in more detail in the context of adversarial robustness in Section~\ref{subsec:llm-limitations}.

Importantly, we assess self-reported awareness during and after tool use, not changes in privacy behavior.
As participants pointed out themselves, raised awareness does not reliably lead to changed behavior. 
Hence, our documentation of how different user profiles experience shifts in awareness provides a basis for future work on behavioral outcomes.

\subsection{\ref{item:rq-usability}: 
How does PRISMe support users' privacy policy-related information-seeking and exploration, and what challenges do users encounter in everyday use?}\label{subsec:disc-rq4}
Participants of all profiles found PRISMe easy and intuitive to use, as evidenced by the interviews and excellent SUS rating. %
Most notably, none required support from a technical person. 
While we did give participants a hands-on tutorial before the study, this suggests that a one-time tutorial after installation suffices.

The layered interface, i.e., an always-visible visual cue, a dashboard, and chat windows for depth, proved effective across all profiles:
\new{\textit{Minimalistic Users}} in particular appreciate the initial visual cues, such as smiley ratings, on which they rely for quick assessments.
\textit{Targeted Explorers} praise our responsive chat providing immediate, detailed responses, while \textit{Novice Explorers} benefit from guided exploration initiated by the overview on the dashboard. 
\textit{Balanced Explorers} profit from the entire range of features offered by PRISMe.
Participants also appreciated customization options across contexts, with some changing the settings both between and within scenarios. 
This allowed them fine-grained control of the level of detail they explore, adjusted to their current context and task at hand.

Some participants wanted easier site-to-site comparisons, noting that assessment criteria varied between websites, and missed weighting of criteria by importance. 
\AF{
This highlights a core challenge of balancing consistent assessment criteria with the wide range of topics users queried: from data collection, sharing, security, and retention to questions about transparency, compliance, profiling, and broader behavioral advice (Section~\ref{subsec:telemetry}). 
Moreover, users might not approach privacy policies with a uniform set of concerns, and relevant questions depend not only on individual preferences but also on the website’s data practices. 
This makes it unlikely that a single, fixed set of criteria can satisfy all users or all websites.
Instead,} for future iterations, we suggest \AF{to utilize the categories of questions participants asked to} derive a small set of standard assessment criteria (e.g., data types, third-party sharing, retention) for consistent comparisons in a side-by-side view, and dynamic, policy-specific set generated by the LLM for contextual nuance. 
Allowing users to furthermore \textit{optionally} select and weight criteria \textit{post-hoc} would satisfy both \textit{Targeted Explorers} (custom settings) and \textit{Minimalistic Users} (default settings).

\AF{A central question for real-world viability is whether and how users would employ PRISMe outside a study context, given that privacy is often a secondary task and many people engage with policies only under necessity, if at all.} 
\textit{Novice Explorers} and \textit{Balanced Explorers} showed the most interest in using the tool in their daily personal lives, describing their experience as \enquote{playful}.
\AF{This opens the door to curiosity-driven checks into the data practices of one's frequently used websites, not necessarily with an intention to switch services, but with the possibility of adjusting one’s behavior on that site. 
Further use cases include: (i) choosing between websites for which users have no strong pre-existing preference, and (ii) higher-stakes contexts, such as financial, medical, or children's data, where deeper engagement is more likely, especially when PRISMe displays a red smiley. 
A source of friction, according to some participants, is the initial assessment runtime (20~s on average, as measured in Section~\ref{subsec:loadingtimes}). 
Since the subsequent chat interaction was generally described as responsive, future engineering work should focus on optimizing the initial assessment. 
Overall, for private use and the majority of users, rather than expecting frequent policy engagement, we see PRISMe as a “background companion” that supports intermittent engagement precisely when motivation spikes. 
}

\textit{Targeted Explorers} in particular voiced their desire to use the tool in a professional setting to query policy specifics.
\AF{They envisioned use cases such as reviewing the privacy policy of their company's website, and comparing it to those of competitors. 
The tool’s exploratory nature could foster privacy literacy in educational settings, offering hands-on instruction regarding the complexities of online data protection.
PRISMe could additionally support third-party risk management by screening potential business partners for alignment with specific data-processing requirements. 
In these settings, privacy is a primary rather than secondary task, and even a 20-second delay for a first assessment is likely tolerable for professionals who prioritize depth and accuracy over speed. 
In turn, the reliability of the RAG’s stricter grounding becomes central, as minimizing hallucinations and ensuring policy fidelity is essential for higher-stakes decision-making. 
}

\subsection{\ref{item:rq-technical}: What types of LLM errors appear in PRISMe’s responses and to what extent does a RAG refinement help mitigation?}\label{subsec:llm-limitations}

Our analysis of the 368 chat responses revealed 6 LLM-related issues: hallucinations, overly euphemistic language, omitting relevant information, generic responses, misunderstanding of queries, and mixed signals.

\textit{Hallucinations} in the output can mislead the user %
and undermine the intended purpose of PRISMe. 
\textit{Minimalistic Users} and \textit{Novice Explorers}, who prioritize efficiency or trust the tool uncritically, are most vulnerable. %
\textit{Balanced Explorers}, more critical of outputs, reflected on responses, and are more likely to spot hallucinations.
\textit{Targeted Explorers}, characterized by skepticism, requested evidence from the policy text as validation and are the least affected. 
We observed that hallucinations were due to queries requiring information that is not part of the privacy policy. 
With our RAG integration, PRISMe filtered out each hallucination and told the user that the policy does not contain sufficient information to answer their question reliably. 
To avoid frustrating users by not answering, PRISMe provides clearly marked generic advice, e.g., recommendations or common practices. %
We also see potential in instructing the LLM to communicate its confidence~\cite{tanneru2024quantifying}. %

\textit{Overly euphemistic or persuasive language} in privacy policies is related to adversarial robustness, because PRISMe is expected to deliver an objective assessment. 
This issue occurring aligns with broader concerns about the LLM-as-a-judge approach~\cite{raina2024llm,chen2024humans}. 
Beyond manipulating responses, the risk of artificially inflating PRISMe’s initial assessment is especially problematic for \textit{Minimalistic users} who mainly rely on the overall assessment without further interaction.
\textit{Targeted} and \textit{Balanced Explorers} showed resilience by critically evaluating responses and identifying overly positive or misleading language, underscoring the need for systems that foster reflective, evidence-based interactions. 

Our RAG successfully removed all occurrences of such affirmative language in all responses. 
Other measures like preprocessing policy text, classifying manipulation potential, refining prompts, or fine-tuning models for more neutral language might also mitigate this issue. 
However, all these approaches including RAG have the potential for 
an arms race between LLM-driven policy assessments and providers’ manipulative tactics to receive better assessments, which mirrors trends in prompt injection~\cite{liu2024formalizing} and ML evasion attacks~\cite{biggio2018wild}.

\textit{Omitting relevant information} is problematic for all user profiles, particularly in listings and enumerations that are too large to be grasped at first glance. 
It may lead to a distorted view of the policy. 
A related issue is giving \textit{generic responses} to specific questions, i.e., providing unnecessarily broad answers. 
Our RAG experiments show a much lesser tendency for both \textit{generic responses} and \textit{omitting information}, enforcing specific answers that are grounded in the privacy policy with successful retrieval. 
While \textit{misunderstanding user queries} did not occur with the RAG, mitigating \textit{mixed signals} involves a design trade-off between fidelity and generic advice, as discussed in Section~\ref{subsec:disc-rq1}.

An engineering task involves further improving the retriever in the RAG-based approach and its embedding model. 
Queries requiring a holistic interpretation of the policy may need to be classified and separately handled, ingesting the full policy without RAG.

\subsection{Further Considerations and Limitations}\label{subsec:limitations}

Our user study provides mainly qualitative evaluation of an LLM-based privacy policy assessment tool, uncovering key interaction patterns and usability challenges. 
While our sample size (N=22) is modest, it aligns with best practices in early-stage usability research, where rich qualitative insights are prioritized over statistical generalizability. 
While questionnaire data supplemented our findings, %
qualitative insights from interviews and in-situ reflections %
offer a nuanced understanding of how users interpret and interact with PRISMe. 
\AF{At the same time,} 
our insights are centered around user perceptions, \VF{and we assessed participants’ privacy familiarity only through self-report, similar to related research~\cite{windl2022automating,zhang2025privcaptcha}.}
\AF{Future work should incorporate both validated privacy literacy measures such as the Online Privacy Literacy Scale (OPLIS)~\cite{masur2017development} and} objective comprehension testing beyond self-reported understanding of privacy policies. %

\AF{Another limitation is a missing comparative baseline against full privacy policies or existing tools. 
Given our exploratory scope, we prioritized understanding how users interpret and interact with PRISMe before conducting comparative studies. 
However, including} efficiency metrics such as measuring how long it takes users to find critical privacy-related information using PRISMe versus other methods \AF{is a key area for further studies. 
In these, one should be careful about requiring participants to read complete policies, as this places substantial time and cognitive demands on them, risking fatigue that could confound the assessment of PRISMe’s usability. 
Meanwhile, conducting comparisons among PRISMe and other tools could contribute to broader community efforts to develop shared evaluation benchmarks and, eventually, a more robust “gold standard” for automated privacy-policy analysis.
}

\AF{Related to this, while we were cautious about validating PRISMe's privacy-policy scoring through careful qualitative checks using crowd-sourced annotations and expert judgment (authors, two external experts, ToS;DR alignment), this is not a substitute for systematic validation. 
What is missing is an established “ground truth” for privacy-policy scoring, which we deem more timely than ever, given the potential of tools such as PRISMe in times of hyper-personalization and micro-targeting~\cite{chouaki2022exploring}.}

Technical limitations include incomplete scraping, which may result in LLM assessments based on partial policies, without alerting users. 
\AF{Importantly, for all predefined websites of the user study, we ensured that the respective policies were retrievable; the only study disruption occurred rarely in the free-exploration scenario.}
The RAG, which we implemented after the user study, solves the alerting issue only for the chat, by stating that PRISMe cannot provide a reliable answer based on the scraped policy.
Adding ML-based techniques for policy classification\VF{~\cite{hosseini2021unifying,bernhard2025multilingual} could further enhance text integrity and improve scraper coverage.} 
Challenges also arise with privacy policies covering multiple services. 
Including HTML code and consent dialogues as LLM inputs could enable step-by-step privacy instructions. 

Further personalizing assessments would require users to directly or indirectly disclose aspects of their identity, indicating a trade-off between personalization of privacy policy assessment and privacy~\cite{freiberger2025explainable}. 
Furthermore, there is a trade-off between privacy and the quality of the explanations for future work to explore: 
running smaller, local LLMs improves privacy around user interests and preferences, but without extensive fine-tuning likely results in reduced output quality compared to a larger model such as GPT-4o.

\section{Conclusion}
\label{sec:conclusion}

In the era of AI-driven hyper-personalization, growing data collection heightens privacy risks, making it increasingly urgent to render complex privacy policies accessible. Existing solutions often fall short of providing understandable, efficient, and trustworthy communication at scale. With PRISMe, we introduced an interactive browser extension that combines layered overviews, customizable exploration, and conversational depth, enabling users to move seamlessly between quick assessments and detailed inquiry.
Our study shows that PRISMe fosters greater privacy awareness and understanding across diverse user profiles, while also surfacing risks such as hallucinations and euphemistic language. By implementing  retrieval-augmented generation into the conversation, we demonstrated both the potential for more trustworthy responses and the trade-offs that arise when constraining or expanding beyond the policy text.
Taken together, our findings highlight the need for tools that are not only technically robust but also carefully attuned to diverse user needs, balancing clarity, fidelity, and breadth. Future research should further investigate how to design privacy policy assistants that remain trustworthy while flexibly supporting different levels of expertise, contexts, and tasks. In doing so, we envision privacy policy exploration tools that go beyond compliance to empower users in making informed decisions in an increasingly data-driven world.

\begin{acks}
The authors acknowledge the financial support by the Federal Ministry of Research, Technology and Space of Germany and by Sächsische Staatsministerium für Wissenschaft, Kultur und Tourismus in the programme Center of Excellence for AI-research „Center for Scalable Data Analytics and Artificial Intelligence Dresden/Leipzig“, project identification number: ScaDS.AI
\end{acks}

\bibliographystyle{ACM-Reference-Format}
\bibliography{literature}

\appendix
\section{Appendix}
\label{sec:appendix}
\subsection{Prompting}\label{app:prompting}
\subsubsection{Prompting Approach for Initial Assessment Generation}
    Your output must be a maximum of 600 words long! You are an expert in data protection and a member of an ethics council. You are given a privacy policy. Your task is to uncover aspects in data protection declarations that are ethically questionable from your perspective. Proceed \textbf{step by step}:
    \begin{enumerate}
        \item \textbf{Criteria:} From your perspective, identify relevant ethical test criteria for this privacy policy as criteria for a later evaluation. When naming the test criteria, stick to standardized terms and concepts that are common in the field of ethics. Keep it short! 
        \item \textbf{Analysis:} Based on this, check for ethical problems or ethically questionable circumstances in the privacy policy. 
        \item \textbf{Evaluation:} Only after you have completed step 2: Rate the privacy policy based on your analysis regarding each of your criteria on a 5-point Likert scale. Explain what this rating means. Explain what the ideal case with 5 points and the worst case with one point would look like. The output in this step should look like this: 
        [Insert rating criterion here]: [insert rating here]/5 [insert line break] 
        [insert justification here]
        
        \item \textbf{Conclusion:} Reflect on your evaluation and check whether it is complete.
    \end{enumerate}
    Important: Check for errors in your analysis and correct them if necessary before the evaluation. You must present your approach clearly and concisely and follow the steps mentioned. Your output must not exceed 600 words.
    
\subsubsection{Prompting Approach for Chat Answer Generation}
\label{sec:chat_prompt}

\textbf{System prompt criteria chat: }Keep it short! Privacy policy: <Privacy policy here> | Rating: <criteria evaluation result here>. Users want to know more about how this rating is justified in the privacy policy. When answering the questions, focus on the given topic of the rating. Keep it short! <Complexity and answer length according to settings>
\\
\textbf{System prompt general chat: }You are an expert in data protection with many years of experience in consumer protection. You have analyzed the following privacy policy and are aware of its risks and ethical implications for users: <Privacy policy here>. 
You should advise users and explain the implications for them in a conversation. <Complexity and answer length according to settings>

\subsubsection{Prompting Approach for our Suggested Question Generation}
\label{sec:suggestion_prompt}

\textbf{System prompt: }Your task is to ask questions about a privacy policy. Your output consists of three questions: 1. question 1; 2. question 2; 3. question 3. Please output the questions in a numbered list. Never repeat questions that have already been asked: <already asked questions here>
\\
\textbf{User prompt: }Specifically: Ask your questions about the privacy policy on the topic: <criterion inserted here>.
Embrace the context of the previous chat: <chat history here>

\subsubsection{Prompting for the RAG experiments}
\label{sec:p_rag}
\textbf{System Prompt: }You are an expert on data protection with many years of experience in consumer protection. You have analyzed the following privacy policy and are aware of its risks and ethical implications for users. Your task is to answer questions based on a privacy policy.\\
\textbf{User Prompt: }Information from the privacy policy is provided as context below:<retrieved context string>. Consider the previous chat history: <context>. Consider the following user preferences: <time>; <complexity>. Respond to the following question based only on the provided context. In case no context is provided, respond with 'The privacy policy does not provide sufficient context to reliably answer this question.' and if general remarks on the question seem helpful, add to the response: 'General remarks: <insert general remarks here>'. If context is provided, utilize it in your response: Question: <query string> Response: 

\subsection{Scenario Description}
\label{sec:scenario}
    We provide you with a browser extension, which you are to use to find out about the privacy practices of websites.
Over the next 20 minutes, you will work through these scenarios on your own. The 20 minutes refers to all scenarios together. If you need help, have a question for us or are stuck, please let us know.

\textbf{Scenario 1:}
You would like to find out about current world events. You regularly read Focus.de and would like to try out their digital offer. 
Find out how Focus.de handles data protection: 
\url{https://www.focus.de/}. 
Take the necessary time to go into detail. 
What is your most important finding?
\\
Consider that you now want to make a digital subscription:
You need to choose a payment method. You are thinking about setting up an account with PayPal. 
Use the tool to find out about PayPal's data protection practices. 
\url{https://www.paypal.com/de}. 
Take the necessary time to go into detail.
What is your most important finding?

\textbf{Scenario 2:}
Imagine you are shopping online. You want to buy books. Searching the web, you come across various web shops: 
\begin{enumerate}

\item \url{https://www.kopp-verlag.de/}

\item \url{https://www.hugendubel.de/}

\item \url{https://www.buchkatalog.de/}

\item \url{https://www.amazon.de/}
\end{enumerate}

Assume that at least the address, contact details, and payment information are required for a purchase. 
For each of these websites, you must consider whether you agree with a purchase and thus the data protection standard of the sites. Use the application to explore the websites with regard to your personal data protection preferences. Please compare the websites. 
Ratings Kopp, Hugendubel, Buchkatalog, Amazon: \\
Where would you most likely buy the book? Why? 

\textbf{Scenario 3: }
Free exploration: visit websites of your choice to find out about their data protection practices. Use the remaining time to explore freely as if you were at home. Let your curiosity run free.

\clearpage

\subsection{Interview Guide}
\label{tab:guide}

\begin{table}[!h]
\small

\begin{tabular}{|p{12.7cm}|p{1.6cm}|}

\hline
\textbf{Exemplary Questions}  & \textbf{Purpose} \\
\hline
What were your first impressions of the extension?\par 
Was there anything surprising or unexpected?\par 
Did you face any issues using it?\par 
How did the information presented by the extension make you feel? &  First thoughts, deviations from expectations  \\
\hline
Did you miss any information being presented in the application?\par 
  How clear was the language used in the application? Were there any terms, phrases, or instructions that you found confusing or unclear?\par Were there moments when you felt overwhelmed by the information presented? \par 
  In case you have specific accessibility needs (e.g. vision or hearing impairment): How well did the extension accommodate this? & Evaluation targeting \ref{item:rq-understandable} \\
\hline
Do you feel like having a better understanding of the issues regarding data protection such a website can have? Explain!\par Did the extension make you aware of any privacy-related issues you were not previously aware of? If yes, can you describe these issues?\par Would you consider changing any of your browsing habits using this extension? If so, how? & Evaluation targeting \ref{item:rq-awareness} \\
\hline
How quickly were you able to find the information of interest using the extension?\par  How effective was the extension to get an overview regarding data protection on the websites?\par
What are your thoughts on the overall design of the extension's interface?\par 
 Were there any features or design elements that you found unnecessary or confusing?\par 
 Is there information the tool missed or did not highlight enough or not in the right presentation style? \par
What changes would you make? \par 
What aspects of the extension contributed most to your satisfaction or dissatisfaction? & Evaluation targeting \ref{item:rq-usability}\\
\hline
Is there anything else you’d like to share about your experience using the extension?  & Open issues  \\
\hline

\end{tabular}

\end{table}

\clearpage

\subsection{Codebook}
\label{sec:codebook}
\begin{table}[!h]
\small
\begin{tabular}{|p{3.5 cm}|p{11.5 cm}|p{1.3 cm}|}
\hline
\textbf{Code}                                                   & \textbf{Description}                                                                                                                                                                     & \textbf{\#Passages} \\ \hline
\multicolumn{3}{|p{14.3 cm}|}{\textbf{Topic 1:} User Attitudes, Motivations, and Behavior
}\\
\hline

indifference                                          & Data protection issues do not concern participants or are of no interest to them                                                                                              & 7                    
\\\hline

emotional response                                    & Tool triggers emotional response of participants                                                                                                                                & 7                    \\ \hline

professional use cases                                & Usecases of the tool that facilitate business processes                                                                                                                         & 10                   \\ \hline

existing privacy   misconceptions                     & Participants having privacy misconceptions or distorted view of reality   regarding privacy                                                                                     & 4                    \\ \hline

multifactorial \&   context-dependent decision making & Participants note that their decisions are based on multiple factors   aside from data protection and depend on the given context                                               & 11\\
\hline

habitualized behavior                                 & Participants are unwilling to change behavior due to existing   habitualized behavior and inconvenience of a change                                                             & 15                   \\ 
\hline

curiosity-driven use                                  & Participants explore using the tool out of curiosity                                                                                                                            & 9                    \\ 
\hline

personal usage interest                               & Participants want to use the tool in their daily personal lives                                                                                                                 & 15                   \\ \hline

insecurity regarding data   protection                & Participants feel insecure about data protection issues                                                                                                                         & 3                    \\ \hline

behavior depends on setting of   use                  & Depending on the given setting like public or own computer use the usage   pattern may differ                                                                                  & 1  \\
\hline

negative predisposition and   distrust                & Participants expect a very low standard of data protection from websites,   are negatively biased by a website's design or have general distrust in   websites' data protection & 30                   \\ \hline

\multicolumn{3}{|p{14.3 cm}|}{\textbf{Topic 2:} Information Quality and Clarity
}\\
\hline

language clarity and   simplicity                     & The used language is easy to understand, clear and simple                                                                                                                       & 35                   \\ \hline

quick and effective overview                          & The tool provides participants with a quick and effective overview of all   relevant information                                                                                & 51                   \\ \hline

Evaluation transparency                               & Aspects regarding how transparent the evaluation process is to users                                                                                                            & 22                   \\ \hline

Levels of detail                                      & Degree of detail and context provided and to what degree it   differentiates evaluations on the different levels of depth in the   application                                  & 35                   \\ \hline

answer quality                                        & Chat answers are helping participants effectively                                                                                                                               & 35                   \\ \hline

Communicated information is   incomplete              & The tool communicates to the user that there is no specific or vague   information on the topic in the policy                                                                   & 7                    \\ \hline

chat flexibility                                      & The chat handles typos, other languages, areas out of context or other   challenges                                                                                             & 6                    \\ \hline

chat consistency                                      & Conversations are consistent and continous in chat, between similar   policies the chat answers for the same questions are also similar                                         & 5                    \\ \hline

less vague \& more to the   point                     & Information presented by the tool should be less vague and more to the   point                                                                                                  & 20                   \\ \hline

adaptability with settings                            & Praise for adjustability of the tool by changing settings                                                                                                                       & 11                   \\ \hline

information rich                                      & The information presented by the application is plentiful, rich and   covering everything relevant                                                                              & 17                   \\ \hline

\multicolumn{3}{|p{14.3 cm}|}{\textbf{Topic 3:} User Experience and Interface Interaction}\\
\hline

good visual cue                                       & The initial smiley icon as visual cue is praised for its design,   placement and increased awareness                                                                            & 29                   \\ \hline

formatting and layout issues                          & Issues addressing the formatting and layout of the application eg. of the   chat output                                                                                         & 36                   \\ \hline

suggestion quality                                    & The quality of the chat query suggestions provided by the tool is praised for its   inspiring and guiding effect but criticised for being too long, unprecise and   not diverse enough     & 23                   \\ \hline

playful                                               & Participants perceive the tool as playful and fun                                                                                                                               & 2                    \\ \hline

visual cue issues                                     & Issues with the visual cue being to intrusive, emotionally loaded,   technichal issues with it due to not appearing, being covered, changing   colors and similar issues        & 22                   \\ \hline

easy and intuitive use                                & How easy and intuitive the application is to use                                                                                                                                & 51                   \\ 
\hline

Loading times                                         & Aspects addressing feedback when the application is loading and long   loading times                                                                                            & 11                   \\ 

\hline

More differentiated initial   scoring                 & More nuanced initial assessment scoring                                                                                                                                         & 4                    \\ 
\hline

Button usability                                      & Whether buttons are easy to use by giving enough feedback when clicked   and being named appropriately                                                                          & 14                   \\ \hline

navigation difficulties                               & Participants face difficulties in navigating the application mostly due   to scrolling                                                                                          & 18                   \\ \hline

challenging to ask precise   questions                & Participants struggle to form precise questions when they want to ask for   specific information                                                                                & 6                    \\ \hline

\end{tabular}
\end{table}
\clearpage
\begin{table}[h!]
\small
\begin{tabular}{|p{3.5 cm}|p{11.5 cm}|p{1.3 cm}|}
\hline
\textbf{Code}                                                   & \textbf{Description}                                                                                                                                                                     & \textbf{\#Passages} \\

                    \hline
criteria dashboard landing   page                     & Put the overview dashboard on the default view                                                                                                                                  & 4                    \\ \hline
                    
                    difficult-to-find or confusing   UI elements          & Including mix up of chat output and text field, terminology explanations   and policy text difficult to find                                                                    & 27                   \\ \hline
                    responsive                                            & The tool responds quickly                                                                                                                                                       & 8                    \\ \hline

accessibility features                                & voice input and audio output                                                                                                                                                    & 7                    \\ \hline

\multicolumn{3}{|p{14.3 cm}|}{\textbf{Topic 4:} Tool Reliability and Trustworthiness}\\
\hline

hallucination risks and LLM   limitations             & Limitations in the LLM's factual accuracy                                                                                                                                       & 27                   \\ \hline

chat relativizes initial   assessment                 & Cases in which chat answers contradict the tool's initial assessment  to some degree                                                                                            & 6                    \\ \hline

trust issue in tool                                   & Participants express issues with trusting the results of the tool                                                                                                               & 17                   \\ \hline

scraper limitations                                   & The scraper used by the tool cannot access some pages privacy policies                                                                                                          & 9                    \\ \hline

inconsistent evaluation   criteria                    & assessment criteria are not fixed and change between policy assessments,   issues with comparibility entail                                                                     & 22                   \\ \hline

validate correct policy                               & The tool should validate whether the correct and full policy has been   scraped                                                                                                 & 13                   \\ \hline

rating accuracy                                       & The rating of a policy by the tool is accurate                                                                                                                                  & 4                    \\ \hline

account for differing   relevance of criteria& Not all assessment criteria are equally relevant, which should be   considered in the assessment                                                                                & 5                    \\ \hline

policy evidence                                       & The tool should utilize quotes or links to the policy as evidence for   presented information                                                                                   & 14                   \\ \hline

\multicolumn{3}{|p{14.3 cm}|}{\textbf{Topic 5:} Recommendations for Features and Functional Improvements}\\
\hline

More customization options                            & Further settings allowing for a more customized use of the tool                                                                                                                 & 4                    \\ \hline

actionable solution                                   & Tool should provide actionable solutions like recommendations,   alternatives, automated adjustment of cookie settings,...                                                      & 13                   \\ \hline

side-by-side comparison                               & Participants wish for side by side comparisons between multiple pages                                                                                                           & 13                   \\ 
 \hline

broader focus on security   threats \& leaks          & The tool should cover cybersecurity more broadly and highlight recent   breaches                                                                                                & 4                    \\ \hline

communicate policy length /   complexity              & Provide background information on policy length and complexity                                                                                                                  & 1                    \\ \hline

contextual feedback privacy   setings                 & Provide feedback on current privacy settings and their context on the   given page                                                                                              & 4                    \\ \hline

read-only variant                                     & Proposed change of the tool to be read-only without interactive chat                                                                                                            & 1                    \\ \hline

multiple services involved                            & Typically not just one service is involved in facilitating the users'   goals and all involved services would need to be checked along the user   journey                       & 1                    \\ \hline

dark mode                                             & Dark mode                                                                                                                                                                       & 1                    \\ \hline

window size \& scaling                                & The window size and scaling of elements in the application should be   bigger and adaptable                                                                                     & 20                   \\ 
\hline

\multicolumn{3}{|p{14.3 cm}|}{\textbf{Topic 6:} Impact on Users}\\
\hline

Pushes reflection process                             & Utilizing the application pushes participants to reflect on data   protection                                                                                                   & 6                    \\ \hline

learning and exploration   process                    & Learning about privacy during exploration                                                                                                                                       & 11                   \\ \hline

improved understanding                                & Participants have learned something about data protection                                                                                                                       & 19                   \\ 
\hline

raised concern                                        & Participants show increased concern regarding data protection due to the   use of the tool                                                                                      & 17                   \\ \hline

improved awareness                                    & The use of the tool made particapants aware of privacy protection issues                                                                                                        & 46                   \\
\hline

\end{tabular}
\end{table}
\clearpage

\end{document}